\documentclass[leqno,authoryear,1p]{article}
\usepackage{times}
\usepackage[german,italian,french,english]{babel}
\usepackage{soul}
\usepackage{amssymb,amsmath}
\usepackage[latin1]{inputenc}
\usepackage{natbib}
\title{The relativity of inertia and reality of nothing}
\author{Alexander Afriat and Ermenegildo Caccese}
\begin{document}
\maketitle
\begin{abstract}
\noindent We first see that the inertia of Newtonian mechanics is absolute and troublesome. General relativity can be viewed as Einstein's attempt to remedy, by making inertia relative, to matter---perhaps imperfectly though, as at least a couple of freedom degrees separate inertia from matter in his theory. We consider ways the relationist (for whom it is of course unwelcome) can try to overcome such undetermination, dismissing it as physically meaningless, especially by insisting on the right transformation properties.
\end{abstract}
\section{Introduction}
The indifference of mechanical phenomena and the classical laws governing them to absolute position, to translation has long been known. This `relativity' extends to the first derivative, velocity, but not to the second, acceleration, which---together with its opposite,\footnote{For motion is inertial when acceleration vanishes. Inertial motion can also be understood, in more Aristotelian terms, as ``natural" rather than ``violent" (accelerated) motion. \citet[p.\ 138]{PdMuN} has a corresponding \emph{dualism between inertia and force}: ``gravity, in the dualism between inertia and force, belongs to inertia, not to force. \emph{In the phenomena of gravitation therefore the inertial- or}, as I prefer to say, \emph{the guidance-field} [\,\dots]." \emph{Cf}.\ \citet[p.~198]{WeylMuK}: ``Dualism between guidance and force." (The translations from German and from French are ours.)\label{Weyl}} \emph{inertia}---has a troubling absoluteness, dealt with in \S\ref{Newton}. General relativity can be seen as Einstein's attempt to overcome that absoluteness (\S\ref{Einstein}), by making inertia relative, to matter. But one can wonder about the extent and nature of the `relativisation.'

Following Einstein we (\S\ref{matter}) take matter to be represented by the energy-momen\-tum tensor\footnote{Indices from the beginning of the Roman alphabet are \emph{abstract indices} specifying valence, contraction (the once-contravariant and twice-covariant trilinear mapping $A^a_{bc}=B^a_{bdc}C^d:\textrm{V}^*\times\textrm{V}\times \textrm{V}\rightarrow\mathbb{R}$, for instance, turns one covector and two vectors into a number) \emph{etc.}, whereas Greek letters are used for space-time coordinate indices running from $0$ to $3$, and $i$, $j$ and $l$ for `spatial' coordinate indices from $1$ to $3$. Sometimes we write $V$ for a four-vector $V^a$, $\alpha$ for a one-form $\alpha_a$, $g$ for the metric $g_{ab}$, and $\langle\alpha ,V\rangle$ for the scalar product $\alpha_aV^a=\langle\alpha_a,V^a\rangle$. The abstract index of the covector $dx^{\mu}=dx^{\mu}_a$, whose valence is obvious, will usually be omitted.} $T^a_b$---rather than by $U^{\mu}_{\nu}=T^{\mu}_{\nu}+t^{\mu}_{\nu}$, which includes the gravitational energy-momentum $t^{\mu}_{\nu}$ whose transformation properties make it too subjective and insubstantial to count. The role of \emph{distant} matter is looked at in \S\ref{distantmatter}. Inertia can be identified with affine or projective structure, as we see in \S\ref{inertia}. In \S\ref{underdetermination} matter appears to underdetermine inertia by ten degrees of freedom, eight of which are (\S\ref{waves}) made to `disappear into the coordinates.' We take such coordinate transformations to be physically meaningless and concentrate on the significance of the remaining quantities instead, which represent the polarisation of gravitational waves.

The physics of gravitational waves seems vulnerable to (admittedly radical) coordinate substitutions, as we see in \S\ref{R&I}: their generation, energy, perhaps even their detection can be `transformed away' if the full range of substitutions, on which general relativity was built, is available. Belief in the production, indeed in the very existence of gravitational radiation is bound up with the binary star PSR 1913$\thinspace+$16, which is considered in  \S\ref{binarystar} and supposed to lose kinetic energy as it spirals inwards; if energy is conserved, the energy lost in one form must be converted, into a perturbation of the surrounding space-time one presumes. But the conservation law is flawed (\S\ref{conservation}), involving, in its integral form, a distant comparison of directions that cannot be both generally covariant and unambiguously integrable. Even the `spiral' behaviour itself, the loss of kinetic energy, and perhaps the oscillation on which detection (\S\ref{Doppler}) is based can be transformed away; as can the energy of the gravitational field, which is customarily assigned using the pseudotensor $t^{\mu}_{\nu}$: while an observer in free fall sees nothing at all, an acceleration would produce energy out of nowhere, out of a mere transformation to another `point of view' or rather state of motion. To take advantage of this fragility of gravitational waves, the relationist wanting motion (inertia in particular) to be `entirely relative' to matter will be mathematically intransigent and attribute physical significance only to notions with the right transformation behaviour---and none to those that can be transformed away---thus allowing him to dispute the reality of the unwelcome freedoms separating matter and inertia, which he can dismiss as mere opinion, as meaningless decoration. If general covariance \emph{has to hold}, matter would seem to determine inertia rather strongly \dots

In the early years of general relativity, Hilbert, Levi-Civita, Schrödinger and others attributed physical meaning only to objects, like tensors, with the right transformation behaviour. Einstein was at first less severe, extending reality to notions with a more radical dependence on the observer's state of motion. With a mathematical argument (\S\ref{defence}) giving a favourable representation of the integral conservation law's transformation properties he persuaded the community to share his tolerance; but would soon, having meanwhile read a manuscript by Cassirer (\S\ref{Cassirer}), change his mind (\S\ref{conversion}) and also require general covariance for physical significance. The relationist can wonder about an argument, and of a widespread indulgence it helped produce, whose proponent and advocate soon adopted the intransigence of his previous opponents.

But rather than as a defence of relationism---for we have no axe to grind---this should be viewed as an exploration of the logical landscape, of certain logical gaps or possibilities the relationist can exploit, especially one that (perhaps unwisely) takes fundamental principles like general covariance more seriously than the lenient pragmatics of day-to-day practice, computation, prediction and success.

The various ways we help or hinder the relationist may sometimes seem arbitrary; to some extent they are arbitrary, or rather influenced by our tastes and interests; but they also take account of the literature and the very full treatments it contains, to which in many cases we have nothing to add.\footnote{For instance we hesitate to help the relationist with the distant masses (which of course constrain inertia) whose influence in the initial-value formulation has been so abundantly considered by Wheeler and others.}

\section{Absolute inertia}\label{background}
\subsection{Newtonian mechanics}\label{Newton}
\noindent Newton distinguished between an ``absolute" space he also called ``true and mathematical," and the ``relative, apparent and vulgar" space in which distances and velocities are physically but imperfectly \emph{measured} down here (rather than exactly divined by the Divinity). Absolute position and motion were not referred to anything. Leibniz identified unnecessary determinations, excess structure\footnote{For a recent treatment see \citet[pp.\ 76-80]{Ryckman}.} in Newton's `absolute' kinematics with celebrated arguments resting on the \emph{principium identitatis indiscernibilium}: as a translation of everything, or an exchange of east and west, produces no observable effect, the situations before and after must be the same, for no difference is discerned. But there were superfluities with respect to Newton's own dynamics too,\footnote{\emph{Cf}.\ \citet[p.\ 178]{Dieks}.} founded as it was on the proportionality of force and acceleration. With his \emph{gran navilio}, \citet[Second day]{Galileo} had already noted the indifference of various \emph{effetti} to inertial transformations; the invariance\footnote{\citet{Newton}, \emph{Corollarium V} (to the laws)} of Newton's laws would more concisely express the indifference of all the \emph{effetti} they governed.\footnote{On this distinction and its significance in relativity see \citet{Dieks}, where the \emph{effetti} are called ``factual states of affairs."}

Modern notation, however anachronistic, can help sharpen interpretation. The derivatives $\dot{\mathbf{x}}={d\mathbf{x}}/{dt}$ and
$\ddot{\mathbf{x}}={d\dot{\mathbf{x}}}/{dt}$ are quotients of differences; already the position difference
\begin{equation*}
\begin{split}
\Delta \mathbf{x}&=\mathbf{x}\textrm{(}t+\varepsilon\textrm{)}-\mathbf{x}\textrm{(}t\textrm{)}\\
&=\mathbf{x}\textrm{(}t+\varepsilon\textrm{)}
+\mathbf{u}-\textrm{[}\mathbf{x}\textrm{(}t\textrm{)}+\mathbf{u}\textrm{]}
\end{split}
\end{equation*}
is indifferent to the addition of a constant $\mathbf{u}$ (which is the same for $\mathbf{x}$ both at $t$ and at $t+\varepsilon$). The velocity
$$\dot{\mathbf{x}}=\lim_{\varepsilon\rightarrow 0}\frac{\Delta\mathbf{x}}{\varepsilon}$$
is therefore unaffected by the three-parameter group $\mathcal{S}$ of translations $\mathbf{x}\mapsto\mathbf{x}+\mathbf{u}$ acting on the three-dimensional space $E$. The difference
\begin{equation*}
\begin{split}
\Delta\dot{\mathbf{x}}&=\dot{\mathbf{x}}\textrm{(}t+\varepsilon\textrm{)}-\dot{\mathbf{x}}\textrm{(}t\textrm{)}\\
&=\dot{\mathbf{x}}\textrm{(}t+\varepsilon\textrm{)}+\mathbf{v}-\textrm{[}\dot{\mathbf{x}}\textrm{(}t\textrm{)}+\mathbf{v}\textrm{]}
\end{split}
\end{equation*}
of velocities is likewise indifferent to the addition of a constant velocity $\mathbf{v}$ (which is the same for $\dot{\mathbf{x}}$ both at $t$ and at $t+\varepsilon$). The acceleration
$$\ddot{\mathbf{x}}=\lim_{\varepsilon\rightarrow 0}\frac{\Delta\dot{\mathbf{x}}}{\varepsilon}$$
is therefore invariant under the six-parameter group $\mathcal{S}\times\mathcal{V}$ which includes, alongside the translations, the group $\mathcal{V}$ of the inertial transformations $\mathbf{x}\mapsto\mathbf{x}+\mathbf{v}t,$ $\dot{\mathbf{x}}\mapsto\dot{\mathbf{x}}+\mathbf{v}$ acting on the space-time $\mathbb{E}=E\times\mathbb{R}$.

The difference
$$\Delta\mathbf{x}=\mathbf{x}\textrm{(}t+\varepsilon\textrm{)}-\mathbf{x}\textrm{(}t\textrm{)}
=\mathbf{x}_a\textrm{(}t+a+\varepsilon\textrm{)}-\mathbf{x}_a\textrm{(}t+a\textrm{)}=\Delta\mathbf{x}_a$$
is also invariant under the group $\mathcal{T}$ of \emph{time translations}. The translation $t\mapsto t+a\in\mathbb{R}$ can be seen as a relabelling of instants which makes $\mathbf{x}$, or rather $\mathbf{x}_a$, assign to $t+a$ the value $\mathbf{x}$ assigned to $t$:  $\mathbf{x}_a\textrm{(}t+a\textrm{)}=\mathbf{x}\textrm{(}t\textrm{)}$. The difference $\Delta\dot{\mathbf{x}}=\Delta\dot{\mathbf{x}}_a$ has the same invariance---as do the quotients ${\Delta\mathbf{x}}/\varepsilon$, ${\Delta\dot{\mathbf{x}}}/\varepsilon$, and the limits $\dot{\mathbf{x}}$, $\ddot{\mathbf{x}}$.

Newton's second law\footnote{``Mutationem motus proportionalem esse vi motrici impressæ, et fieri secundum lineam rectam qua vis illa imprimitur."} is `covariant' with respect to the group $\mathcal{R}=\textrm{SO(}\mathcal{S}\textrm{)}$ of rotations $R:\thinspace E\rightarrow E$, which turn the  ``straight line along which the force is applied" with the ``change of motion," in the sense that the two rotations $F\mapsto RF,\,\ddot{\mathbf{x}}\mapsto R\ddot{\mathbf{x}}$, taken together, maintain the proportionality of force and acceleration expressed by the law: $[F\thicksim\ddot{\mathbf{x}}]\Leftrightarrow[RF\thicksim R\ddot{\mathbf{x}}].$ We can say the second law is indifferent\footnote{For Newton's forces are superpositions of fundamental forces $F=f(|\mathbf{x}_2-\mathbf{x}_1|,|\dot{\mathbf{x}}_2-\dot{\mathbf{x}}_1|,|\ddot{\mathbf{x}}_2-\ddot{\mathbf{x}}_1|,\dots)$, covariant under $\mathcal{G}$, exchanged by pairs of points.} to the action of the ten-parameter Galilei group\footnote{See \citet[pp.\ 224-9]{LL}.} $\mathcal{G}=(\mathcal{S}\times\mathcal{V})\rtimes(\mathcal{T}\times\mathcal{R})$ with composition
$$(\mathbf{u},\mathbf{v},a,R)\rtimes(\mathbf{u}',\mathbf{v}',a',R')=(\mathbf{u}+R\mathbf{u}'+a'\mathbf{v},\mathbf{v}+R\mathbf{v}',a+a',RR')\textrm{,}$$
$\rtimes$ being the semidirect product. So the \emph{absolute} features of Newtonian mechanics---acceleration, force, inertia, the laws---emerge as invariants of the Galilei group, whose transformations change the \emph{relative} ones: position, velocity and so forth. A larger group admitting acceleration would undermine the laws, requiring generalisation with other forces.

\citet{Cartan}\footnote{See also \citet[\S III]{Friedman}, \citet[\S17.5]{Penrose}.} undertook such a generalisation, with a larger group, new laws and other forces. The general covariance of his Newtonian formalism (with a flat connection) may seem to make inertia and acceleration relative, but in fact the meaningful acceleration in his theory is not ${d^2x^{i}}/{dt^2}$, which can be called \emph{relative}\footnote{In \citet{Baker} there appears to be a confusion of the two accelerations as they arise---in much the same way---in general relativity. The acceleration ${d^2x^{\mu}}/{d{\tau}^2}\neq 0$ Baker sees as evidence of the causal powers possessed by an ostensibly empty space-time with $\Lambda\neq 0$ is merely \emph{relative}; even with $\Lambda\neq 0$ free bodies describe geodesics, which are wordlines whose \emph{absolute} acceleration vanishes. The sensitivity of projective or affine structure to the cosmological constant $\Lambda$ would seem to be more meaningful, and can serve to indicate similar causal powers.} (to the coordinates), but the \emph{absolute}
\begin{equation}\label{Cartan}
A^i=\frac{d^2x^i}{dt^2}+\sum^3_{j,l=1}\mathit{\Gamma}^i_{jl}\frac{dx^j}{dt}\frac{dx^l}{dt}
\end{equation}
($i=1,2,3$ and the time $t$ is absolute). Relative acceleration comes and goes as coordinates change, whereas absolute acceleration is generally covariant and transforms as a tensor: if it vanishes in one system it always will. The two accelerations coincide with respect to inertial coordinates, which make the connection components\footnote{The abstract index representing the valence of the `partial derivation' vector $\partial_i=\partial^a_i={\partial}/{\partial x^i}$ tangent to the $i$th coordinate line will be omitted.} $\mathit{\Gamma}^i_{jl}=\langle dx^i,\nabla_{\partial_j}\partial_l\rangle$ vanish. The absolute acceleration of inertial motion vanishes however it is represented---the connection being there to cancel the acceleration of non-inertial coordinates.

So far, then, we have two formal criteria of inertial motion:
\begin{itemize}
\item $\ddot{\mathbf{x}}=\mathbf{0}$ in Newton's theory
\item $A^i=0$ in Cartan's.
\end{itemize}
But Newton's criterion doesn't really get us anywhere, as the vanishing acceleration has to be referred to an inertial frame in the first place; to Cartan's we are about to return.

\citet[p.\ 770; 1988, p.\ 40; 1990, p.~28]{Einstein1916} and others have appealed to the \emph{simplicity of laws} to tell inertia apart from acceleration: inertial systems admit the simplest laws. Condition
$\ddot{\mathbf{x}}=\mathbf{0}$, for instance, is simpler than
$\ddot{\mathbf{y}}+\mathbf{a}=\mathbf{0}$, with a term $\mathbf{a}$ to compensate the acceleration of system $y$. But we have just seen that Cartan's theory takes account of possibile acceleration \emph{ab initio}, thus preempting subsequent complication---for accelerated coordinates do not appear to affect the syntactical form\footnote{\emph{Cf}.\ \citet[p.\ 186]{Dieks}.} of (\ref{Cartan}), which is complicated to begin with by the connection term. One could argue that the law simplifies when that term disappears, when the coefficients $\mathit{\Gamma}^i_{jl}$ all vanish; but then we're back to the Newtonian condition $\ddot{\mathbf{x}}=\mathbf{0}$. And just as that condition requires an inertial system in the first place, Cartan's condition $A^i=0$ requires a connection, which is pretty much equivalent: it can be seen as a convention stipulating how the three-dimensional simultaneity surfaces are `stitched'\footnote{See \citet[\S\S1,2]{Earman}, for instance.} together by a congruence of (mathematically) arbitrary curves \emph{defined} as geodesics. The connection would then be determined, \emph{a posteriori} as it were, by the requirement that its coefficients vanish for those inertial curves. Once one congruence is chosen the connection, thus determined, provides all other congruences that are inertial with respect to the first. So the initial geodesics, by stitching together the simultaneity spaces, first provide a notion of rest and velocity, then a connection, representing inertia and acceleration. The Newtonian condition $\ddot{\mathbf{x}}=\mathbf{0}$ presupposes the very class of inertial systems given by the congruence and connection in Cartan's theory. So we seem to be going around in circles: \emph{motion is inertial if it is inertial with respect to inertial motion}.

We should not be too surprised that purely formal criteria are of little use on their own for the identification of something as physical as inertia. But are more physical, empirical criteria not available? Suppose we view Newton's first law, his `principle of inertia,' as a special case of the second law $F=m\ddot{\mathbf{x}}$ with vanishing force (and hence acceleration). So far we have been concentrating on the more mathematical right-hand side, on vanishing acceleration; but there is also the more physical left-hand side $F=\mathbf{0}$: can inertial systems not be characterised\footnote{\citet[p.\ 772; 1988, p.~40; 1990, p.~59]{Einstein1916}} as free and far from everything else? Even if certain bodies may be isolated enough to be almost entirely uninfluenced by others, the matter remains problematic. For one thing we have no direct access to such roughly free bodies, everything around us gets pulled and accelerated. And the absence of gravitational force is best assessed with respect to an inertial system,\footnote{An anonymous referee has pointed out that inertial systems can be large and rigid in flat space-times, but not with curvature; where present, tidal effects prevent inertial motion from being rigid, and even rule out large inertial frames; but see \S\ref{midwife}.} which is what we were after in the first place.

Just as the \emph{absence} of force has been appealed to for the identification of inertia, its \emph{presence} can be noted in an attempt to characterise acceleration; various passages\footnote{``Distinguuntur autem quies et motus absoluti et relativi ab invicem per proprietates suas et causas et effectus"; ``Causæ, quibus motus veri et relativi distinguuntur ab invicem, sunt vires in corpora impressæ ad motum generandum"; ``Effectus, quibus motus absoluti et relativi distinguuntur ab invicem, sunt vires recedendi ab axe motus circularis"; ``Motus autem veros ex eorum causis, effectibus, et apparentibus differentiis colligere, et contra ex motibus seu veris seu apparentibus eorum causas et effectus, docebitur fusius in sequentibus."} in the \emph{scho\-lium} on absolute space and time show that Newton, for instance, proposed to tell apart inertia and acceleration through \emph{causes}, \emph{effects}, \emph{forces}.\footnote{\emph{Cf}.\ \citet{Rynasiewicz2}.} In the two experiments described at the end of the \emph{scholium}, involving the bucket and the rotating globes, there is an interplay of local causes and effects: the rotation of the water causes it to rise on the outside; the forces applied to opposite sides of the globes cause the tension in the string joining them to vary. But this doesn't get us very far either; our problem remains, as we see using the distinction drawn above between absolute acceleration $A^i$ and relative acceleration $d^2x^i/dt^2$, which surprisingly corresponds to a distinction Newton himself is groping for in the following passage from the \emph{scholium}:
\begin{quote}
The causes by which true and relative motions are distinguished, one from the other, are the forces impressed upon bodies to generate motion. True motion is neither generated nor altered, but by some force impressed upon the body moved; but relative motion may be generated or altered without any force impressed upon the body. For it is sufficient only to impress some force on other bodies with which the former is compared, that by their giving way, that relation may be changed, in which the relative rest or motion of this other body did consist. Again, true motion suffers always some change from any force impressed upon the moving body; but relative motion does not necessarily undergo any change by such forces. For if the same forces are likewise impressed on those other bodies, with which the comparison is made, that the relative position may be preserved, then that condition will be preserved in which the relative motion consists. And therefore any relative motion may be changed when the true motion remains unaltered, and the relative may be preserved when the true suffers some change. Thus, true motion by no means consists in such relations.\footnote{``Causæ, quibus motus veri et relativi distinguuntur ab invicem, sunt vires in corpora impressæ ad motum generandum. Motus verus nec generatur nec mutatur nisi per vires in ipsum corpus motum impressas: at motus relativus generari et mutari potest absq; viribus impressis in hoc corpus. Sufficit enim ut imprimantur in alia solum corpora ad quæ fit relatio, ut ijs cedentibus mutetur relatio illa in qua hujus quies vel motus relativus consistit. Rursus motus verus a viribus in corpus motum impressis semper mutatur, at motus relativus ab his viribus non mutatur necessario. Nam si eædem vires in alia etiam corpora, ad quæ fit relatio, sic imprimantur ut situs relativus conservetur, conservabitur relatio in qua motus relativus consistit. Mutari igitur potest motus omnis relativus ubi verus conservatur, et conservari ubi verus mutatur; et propterea motus verus in ejusmodi relationibus minime consistit."}
\end{quote}
The absolute \emph{motus} of a body $\beta$ requires a force on $\beta$, but to produce relative \emph{motus} the force can act on the reference body $\gamma$ instead; and relative \emph{motus} can even be cancelled if force is applied to both $\beta$ and $\gamma$. The translators, Motte and Cajori, render \emph{motus} as ``motion" throughout, but the passage only makes sense (today) if we use \emph{acceleration}, for most occurrences at any rate: Newton first speaks explicitly of the generation or alteration of motion, to establish that `acceleration' is at issue; having settled that he abbreviates and just writes \emph{motus}---while continuing to mean acceleration. And he distinguishes between a true acceleration and a relative acceleration which can be consistently interpreted, however anachronistically, as $A^i$ and $d^2x^i/dt^2$. Of course Newton knows neither about connections nor affine structure, nor even matrices; but he is clearly groping for something neither he nor we can really pin down using the mathematical resources then available. It may not be pointless to think of a `Cauchy convergence' of sorts towards something which at the time is unidentified and alien, and only much later gets discovered and identified as the goal towards which the intentions, the gropings were tending.

When Newton states, in the second law, that the \emph{mutationem motus} is proportional to force, he could mean either the true acceleration or the relative acceleration; indeed it is in the spirit of the passage just quoted to distinguish correspondingly---pursuing our anachronism---between a \emph{true force} $F^i=mA^i$ and a \emph{relative force} $f^i=m\thinspace d^2x^i/dt^2$. This last equation represents one condition for two unknowns, of which one can be fixed or measured to yield the other. But the relative force $f^i$ is the wrong one. The `default values' for both force and acceleration, the ones Newton is really interested in, the ones he means when he doesn't specify, the ones that work in his laws, are the `true' ones: true force and true acceleration. And even if $F^i=mA^i$ also looks like one condition for two unknowns, the true acceleration $A^i$ in fact conceals \emph{two} unknowns, the relative acceleration $d^2x^i/dt^2$ and the difference $A^i-d^2x^i/dt^2$ representing the absolute acceleration of the coordinate system.\footnote{All sorts of questions can be raised about the direct measurability of the true force.} Nothing doing then, we're still going around in circles: the inertia of Newtonian mechanics remains absolute, and cannot even be `made relative' to force.

But what's wrong with absolute inertia? In fact it can also be seen as `relative,' but to something---mathematical structure or the \emph{sensorium Dei} or absolute space---that isn't really there, that's too tenuous, invisible, mathematical, ætherial, unmeasurable or theological to count as a cause, as a physically effective circumstance, for most empiricists at any rate.\footnote{\emph{Cf}.\ \citet[pp.~771-2; 1917b, p.~49; 1990, p.~57]{Einstein1916}, \citet[pp.\ 31, 38, 39]{Cassirer}, \citet[\S2.2.2]{Rovelli}.\label{invisible}} The three unknowns of $F^i=mA^i$ are a problem because in Newtonian mechanics affine structure, which determines $A^i-d^2x^i/dt^2$, is unobservable. By relating it to matter Einstein would give inertia a solid, tangible, empirically satisfactory foundation.
\subsection{Einstein}\label{Einstein}
\noindent General relativity can be seen as a response to various things. It suits our purposes to view it as a reaction to two `absolute' features of Newtonian mechanics, of Newtonian inertia, to which Einstein objected: i.~an observable effect arising out of an unobservable cause; and ii.~action without passion. In \S\ref{RoI} we will wonder how complete a response it would prove.

i.~We have just seen that Newton proposed to find absolute acceleration through its causes and effects. Einstein also speaks of cause and effect---and practically seems to be addressing Newton and his efforts to sort out absolute and relative \emph{motus}---in his exposition of the thought experiment at the beginning of ``Die Grundlage der allgemeinen Relativitätstheorie" \citeyearpar[p.~771]{Einstein1916}. There he brings together elements of Newton's two experiments---rotating fluid, two rotating bodies: Two fluid bodies of the same size and kind, $S_1$ and $S_2$, spin with respect to one another around the axis joining them while they float freely in space, far from everything else and at a considerable, unchanging distance from each other. Whereas $S_1$ is a sphere $S_2$ is ellipsoidal. Einstein's analysis of the difference betrays positivist zeal and impatience with metaphysics. Newton, who could be metaphysically indulgent to a point of mysticism, might---untroubled by the absence of a manifest local cause---have been happy to view the deformation of $S_2$ as the effect of an absolute rotation it would thus serve to reveal. Einstein's epistemological severity makes him more demanding; he wants the observable cause\footnote{\citet[p.\ 771]{Einstein1916}; \emph{cf}.\ footnote \ref{invisible} above. Einstein wants visible effects to have visible causes; \emph{cf}.\ \citet[pp.~64-94]{Poincare1908}, who sees ``chance" when ``large" effects have ``small" causes---which can even be too small to be observable; and \citet[p.~162]{Russell1961}: ``a very small force might produce a very large effect. [\,\dots] An act of volition may lead one atom to this choice rather than that, which may upset some very delicate balance and so produce a large-scale result, such as saying one thing rather than another."} of the differing shapes; seeing no \emph{local} cause, within the system, he feels obliged to look elsewhere and finds an \emph{external} one in distant masses which rotate with respect to $S_2$.

ii.~\citet{Einstein1922} also objects to ``the postulation," in Newtonian mechanics, ``of a thing (the space-time continuum) which acts without being acted upon."\footnote{P.~58: ``Erstens nämlich widerstrebt es dem wissenschaftlichen Verstande, ein Ding zu setzen (nämlich das zeiträumliche Kontinuum), das zwar wirkt, auf welches aber nicht gewirkt werden kann." \emph{Cf}.\ \citet[p.~51]{Weyl1931}: ``Space accordingly acts on [things], the way one necessarily conceives the behaviour of an absolute God on the world: the world subject to his action, he spared however of any reaction."} Newtonian space-time structure---inertial structure in particular---has a lopsided, unreciprocated relationship with matter, which despite being guided by it does nothing in return.

General relativity responds to absolute inertia by relating inertia to \emph{matter}, which has a more obvious physical presence than mathematical background structure or the \emph{sensorium Dei}. In ``Prinzipielles zur allgemeinen Relativitätstheorie" \citeyearpar{Einstein1918} Einstein goes so far as to claim that inertia\footnote{In fact he speaks of the ``$G$-field" \citeyearpar[p.~241]{Einstein1918}, ``the state of space described by the fundamental tensor," by which inertia is represented: ``Inertia and weight are essentially the same. From this, and from the results of the special theory of relativity, it follows necessarily that the symmetrical `fundamental tensor' ($g_{\mu\nu}$) determines the metrical properties of space, the inertial behaviour of bodies in it, as well as gravitational effects."} in his theory is entirely determined\footnote{\emph{Ibid}.\ p.~241: ``\emph{Mach's principle:} The $G$-field is \emph{completely} determined by the masses of bodies." See \citet{Hoefer} on ``Einstein's formulations of Mach's principle."\label{MachschesPrinzip}} by matter, which he uses $T_{ab}$ to represent:
\begin{quote}
Since mass and energy are the same according to special relativity, and energy is formally described by the symmetric tensor ($T_{\mu\nu}$), the $G$-field is determined by the energy tensor of matter.\footnote{\emph{Ibid}.\ 241-2: ``Da Masse und Energie nach den Ergebnissen der speziellen Relativitätstheorie das Gleiche sind, und die Energie formal durch den symmetrische Energietensor ($T_{\mu\nu}$) beschrieben wird, so besagt dies, daß das $G$-Feld durch den Energietensor der Materie bedingt und bestimmt sei."}
\end{quote}
He explains in a footnote (p.~241) that this \emph{Machsches Prinzip} is a generalisation of Mach's requirement \citeyearpar[\S2.6]{Mach} that inertia be derivable from interactions between bodies.\footnote{\citet{Barbour} is full of excellent accounts; see also \citet[pp.\ 105-8]{Earman}, \citet{Mamone} and \citet[\S2.4.1]{Rovelli}.}

So we seem to be wondering about what Einstein calls \emph{Mach's principle}, which provides a convenient label, and is something along the lines of \emph{matter determines inertia}. We have seen what a nuisance absolute inertia can be; to remedy Einstein made it relative, to matter; we accordingly consider the extent and character of his `relativisation,' of the \emph{determination of inertia by matter}.

\section{The relativity of inertia}\label{RoI}
\subsection{Matter}\label{matter}
\noindent To begin with, what is matter? \citet{Einstein1918}, we have seen, used $T^a_b$ to characterise it, but maybe one should be more permissive and countenance less substantial stuff as well. Einstein proposed
\begin{equation}\label{pseudotensor}
t^{\mu}_{\nu}=\frac{1}{2}\delta^{\mu}_{\nu}g^{\sigma\tau}\mathit{\Gamma}^{\lambda}_{\sigma\rho}\mathit{\Gamma}^{\rho}_{\tau\lambda}-g^{\sigma\tau}\mathit{\Gamma}^{\mu}_{\sigma\rho}\mathit{\Gamma}^{\rho}_{\tau\nu}
\end{equation}
for the representation of \emph{gravitational} mass-energy; matter without mass, or mass away from matter, are hard to imagine; so perhaps we can speak of gravitational \emph{matter}-mass-energy.\footnote{\emph{Cf}.\ \citet[p.\ 82]{Russell1927}: ``We do not regard energy as a ``thing," because it is not connected with the qualitative continuity of common-sense objects: it may appear as light or heat or sound or what not. But now that energy and mass have turned out to be identical, our refusal to regard energy as a ``thing" should incline us to the view that what possesses mass need not be a ``thing.""} How about $U^{\mu}_{\nu}=T^{\mu}_{\nu}+t^{\mu}_{\nu}$ then, rather than just $T^a_b$? Several drawbacks come to mind. The right-hand side\footnote{Its convenient form is assumed with respect to coordinates satisfying $1=\sqrt{-\mathbf{g}}$, where $\mathbf{g}$ is the determinant of the metric.} of (\ref{pseudotensor}) shows how such `matter' would be related to the notoriously untensorial connection components. In free fall, when they vanish, the pseudotensor $t^{\mu}_{\nu}$ does too,\footnote{Issues related to the domain of \emph{Wegtransformierbarkeit} are considered in \S\ref{midwife}. \emph{Wegtransformierbarkeit} or `away-transformability' is a useful notion for which there seems to be no English word.} which means that gravitational matter-mass-energy would be a \emph{matter of opinion},\footnote{\emph{Cf}.\ \citet[p.\ 519]{EarmanNorton}.} its presence depending on the state of motion of the observer. The distribution of matter-mass-energy in apparently empty space-time would accordingly depend on the choice of coordinates. To be extremely liberal one could even fill the whole universe, however empty or flat, on grounds that matter-mass-energy is potentially present everywhere, as an appropriate acceleration could produce it anywhere.

A superabundance of matter would help constrain inertia and hence make `Mach's principle'---indeed any relationist claim or principle---easier to satisfy, perhaps to a point of vacuity. The relationist would also be brought uncomfortably close to his `absolutist' opponent, who believes there is more to inertia than one may think, that it goes beyond and somehow transcends determination by matter. Indeed if we spread matter too liberally we hardly leave the relationist and absolutist much room to differ. \citet{Rynasiewicz} has already dismissed their debate as ``outmoded"; the surest way to hasten its complete (and regrettable) demise is to impose agreement, by a questionable appeal to a dubious object which can cover the universe with slippery coordinate-dependent matter that disappears in free fall and reappears under acceleration.\footnote{This is no peculiarity of general relativity, as an anonymous referee has pointed out: even in older theories the local energy density can disappear and reappear under coordinate changes.} We began with Newton, Leibniz and Galileo, have been guided by a continuity connecting their preoccupations with Einstein's, and accordingly adopt a notion of matter that differs as little as possible (within general relativity) from theirs: hence $T^a_b$, rather than the ill-behaved $U^{\mu}_{\nu}=T^{\mu}_{\nu}+t^{\mu}_{\nu}$.

\subsection{Distant matter}\label{distantmatter}
\noindent This paper is much more about general relativity than about Mach himself; it is certainly not about Mach's own formulations of his principles. The vagueness and ambiguities of \citet[\S2.6]{Mach} have given rise to an abundance of `Mach's principles,' many of which are represented in \citet{Barbour}. Mach and \citet[p.~772]{Einstein1916} both speak of ``distant" matter, which indeed figures in several versions of `Mach's principle': one can say it is part of the `Machian tradition,' conspicuously associated with Einstein, Wheeler, Barbour and others. But distant matter can affect inertia in two very different ways: i.~the `deceptive continuity' or `average character' of $\rho$; and ii.~`field-theoretical holism.'

i.~Einstein's equation $G_{ab}\textrm{(}x\textrm{)}=T_{ab}\textrm{(}x\textrm{)}$ seems to express a circumscribed (direct) relationship between inertia and matter at (or around) point $x$. The matter-energy-momentum tensor
$$T^{ab}\textrm{(}x\textrm{)}=\rho\textrm{(}x\textrm{)}V^aV^b\textrm{,}$$
for instance, describing a dust with density $\rho$ and four-velocity $V^a$, would (directly) constrain inertia at $x$, not at other points far away. But much as in electromagnetism, the `continuity' of $\rho$ is deceptive. Once the scale is large enough to give a semblance of continuity to the density $\rho$, almost all the celestial bodies contributing to the determination of $\rho\textrm{(}x\textrm{)}$ will be very far, on any familiar scale, from $x$. \citet{KosmologischeBetrachtungen} sees $\rho$ as an average, and speaks of `spreading':
\begin{quote}
The metrical structure of this continuum must therefore, as the distribution of matter is not uniform, necessarily be most complicated. But if we are only interested in the structure in the large, we ought to represent matter as evenly spread over enormous spaces, so that its density of distribution will be a function that varies very slowly.\footnote{P.~135: ``Die metrische Struktur dieses Kontinuums muß daher wegen der Ungleichmäßigkeit der Verteilung der Materie notwendig eine äußerst verwickelte sein. Wenn es uns aber nur auf die Struktur im großen ankommt, dürfen wir uns die Materie als über ungeheure Räume gleichmäßig ausgebreitet vorstellen, so daß deren Verteilungsdichte eine ungeheuer langsam veränderliche Funktion wird."}
\end{quote}
Needless to say, all the matter involved in the determination of $\rho\textrm{(}x\textrm{)}$ will be very close to $x$ on the largest scales; but matter far from $x$ even on those scales has a role too, a field-theoretical role, as we shall now see.

ii.~\citet{Riemann} considered the possibility of a discrete manifold $D$, with denumerable elements $D_1,D_2,\dots$. Of course the value $\varphi_r=\varphi\textrm{(}D_r\textrm{)}$ of a (scalar) field $\varphi$ at $D_r$ will be completely unconstrained by the values $\varphi_s$ at other points $D_s$ if no restrictions are imposed. On its own the `boundary condition' $\varphi_s\rightarrow 0$ \emph{as} $s\rightarrow\infty$---or even the stronger condition $\varphi_s$ \emph{vanishes for} $s>1$---will not constrain $\varphi_1$ at all. But the further requirement that, say, $$|\varphi_r-\varphi_s|<\frac{1}{2}\min\{|\varphi_r|,|\varphi_s|\}$$
for adjacent points (\emph{i.e}.\ $|r-s|=1$) gives, by heavily constraining either value once the other is fixed, the crudest idea of how boundary conditions act.

Of course the manifolds involved in general relativity are continuous, with smooth fields on them, which leads to subtler, less trivial constraint: such fields can undulate, propagate perturbations, drag and so forth; the constrained relationship between neighbouring values can ripple across the universe at the speed of light. The value $\mathbf{R}\textrm{(}x\textrm{)}$ of a field $\mathbf{R}$ at point $x$ can be indirectly constrained through restrictions imposed by another field $\mathbf{T}$ on the values $\mathbf{R}\textrm{(}x'\textrm{)}$ at points $x'$ far away; or directly, by the physicist, who may require for instance that $\mathbf{R}$ itself vanish somewhere---here one speaks of `boundary conditions.' If the universe foliates into spatially non-compact simultaneity surfaces, such boundary conditions have to be imposed, typically asymptotic flatness. But this, wrote \citet{KosmologischeBetrachtungen}, is at odds with the relativity of inertia: ``inertia would be \emph{influenced} but not \emph{determined} by matter"\footnote{P.~135: ``Somit würde die Trägheit durch die (im Endlichen vorhandene) Materie zwar \emph{beeinflußt} aber nicht \emph{bedingt}."}---since the full determination requires the `additional,' physically `extraneous' stipulation of boundary conditions. So he did away with boundary conditions by doing away with the boundary: he proposed a universe foliating into spatially compact simultaneity surfaces (without boundary), which lend themselves to `global' Machian interpretations by favouring the determination of inertia by matter.\footnote{Both kinds of foliation have received ample attention in the literature; see \citet{Wheeler1959}, \citet{CB}, \citet{OMY}, \citet{IsenbergWheeler}, \citet{CBY}, \citet{Isenberg}, \citet{York}, \citet[\S5]{CiufoliniWheeler}, \citet{LusannaPauri, LusannaPauri2, LusannaPauri3}, \citet{Lusanna}, \citet{LusannaAlba}.}

Even if the determination is partly field-theoretical, holistic, global, non-local,\footnote{\citet[pp.~719-20 for instance]{LusannaPauri} consider such non-locality in the Hamiltonian formulation.} \emph{we will concentrate on the `punctual' determination, on the arithmetic and comparison of freedom degrees at a point}. Words like ``determination," ``over/underdetermination" or ``freedom" are often referred to a single point---by Einstein and others---even in field-theoretical contexts (where more holistic influences are also at work), and seem neither illegitimate, meaningless nor inappropriate when applied so locally.\footnote{Specification of circumstances at a point is not enough, as an anonymous referee has pointed out, for \emph{prediction} in a field theory, where much more (Cauchy data on a Cauchy surface) would have to be indicated.}

It is worth mentioning that Einstein's own position on the matter of punctual rather than field-theoretical, non-local determination is confusing. In ``Kosmologische Betrachtungen zur allgemeinen Relativitätstheorie" \citeyearpar{KosmologischeBetrachtungen}, which is all about field\-theoretical holism, he writes:
\begin{quote}
According to general relativity, the metrical character (curvature) of the four-dimensional space-time continuum is determined at every point by the matter that's there, together with its state.\footnote{P.~135: ``Der metrische Charakter (Krümmung) des vierdimensionalen raumzeitlichen Kontinuums wird nach der allgemeinen Relativitätstheorie in jedem Punkte durch die daselbst befindliche Materie und deren Zustand bestimmt."}
\end{quote}
And he often counts degrees freedom at a point, saying that one object there over- or under-determines another.

\subsection{Inertia}\label{inertia}
\noindent Inertial motion is free and not \emph{force}d by alien influences to deviate from its natural course. The characterisation is general, its terms take on specific meaning in particular theories: in general relativity, inertial motion is subject only to gravity and not to electromagnetic or other forces; we accordingly identify \emph{inertia} with the structures that guide the free fall of small\footnote{We only know that \emph{test} bodies follow geodesics, as an anonymous reviewer has emphasised. Bodies large enough to influence projective structure may be guided by it in a different way: ``Since we do not know how to solve Einstein's equations with matter, we do not know whether `dynamical masses' follow  geodesics."} bodies (perhaps the hands of clocks too) by determining the (possibly parametrised) geodesics they describe.\footnote{\emph{Cf}.\ \citet{Dorato}.}

We have seen that Einstein identifies inertia with the metric $g$, which in general relativity---where $\nabla g$ vanishes (along with torsion)---corresponds to the affine structure given by the Levi-Civita connection $\nabla=\Pi_\mathbf{0}$, with twenty degrees of freedom. It gives the \emph{parametrised} geodesics
$\sigma_{\mathbf{0}}:\thinspace \textrm{(}a_\mathbf{0},b_\mathbf{0}\textrm{)}\rightarrow M;\thinspace s_\mathbf{0}\mapsto\sigma_\mathbf{0}\textrm{(}s_\mathbf{0}\textrm{)}$
through $\nabla_{\dot{\sigma}_\mathbf{0}}\dot{\sigma}_\mathbf{0}=\mathbf{0}$, and represents the `inertia' of the parameter, hence of the hands of clocks, along with that of matter. ($M$ is the differential manifold representing the universe.)

But time and clocks may be less the point here than plain free fall. Weyl\footnote{See footnote \ref{Weyl}, and \citet{Weyl}; or \citet[p.~233]{Malament} for a more modern treatment.} identified inertia with the weaker \emph{projective} structure $\Pi$, which gives the `generalised geodesics'\footnote{Or alternatively the unparametrised geodesics, in other words just the image $\mathfrak{I}\textrm{(}\sigma\textrm{)}=\mathfrak{I}\textrm{(}\sigma_{\mathbf{0}}\textrm{)}\subset M$.}
$\sigma:\textrm{(}a,b\textrm{)}\rightarrow M;\thinspace s\mapsto\sigma\textrm{(}s\textrm{),}$
through $\nabla_{\dot{\sigma}}\dot{\sigma}=\lambda\dot{\sigma}$.
Projective structure just represents free fall, in other words the inertia of bodies alone, not of bodies \emph{and} the hands of accompanying clocks. One can say it is purely `material,' rather than `materio-temporal.'

In the class $\Pi=\{\Pi_{\alpha}:\alpha\in\Lambda_1\textrm{(}M\textrm{)}\}$ of connections projectively equivalent to $\nabla$, a particular connection $\Pi_{\alpha}$ is singled out by a one-form $\alpha$, which fixes the parametrisations $s$ of all the generalised geodesics $\sigma$. So projective structure has twenty-four degrees of freedom, four---namely $\alpha_0,\dots,\alpha_3$---more than affine structure; $\alpha_{\mu}=\langle\alpha,\partial_{\mu}\rangle$. We can write
$$\langle dx^{\mu},\Pi_{\alpha\thinspace\partial_{\nu}}\partial_{\kappa}\rangle
=\mathit{\Gamma}^{\mu}_{\nu\kappa}+\delta^{\mu}_{\nu}\alpha_{\kappa}+\delta^{\mu}_{\kappa}\alpha_{\nu}\textrm{,}$$
where the $\mathit{\Gamma}^{\mu}_{\nu\kappa}$ are the components of the Levi-Civita connection. The most meaningful part of the added freedom appears to be the `acceleration'
$$\lambda=-2\langle\alpha,\dot{\sigma}\rangle=-2\,\alpha_{\mu}\frac{d\sigma^{\mu}}{ds}=-\left(\frac{ds}{ds_{\mathbf{0}}}\right)^{\negthinspace\negthinspace 2}\frac{d^2s_{\mathbf{0}}}{ds^2}$$
of the parameter $s$ along the generalised geodesic $\sigma$ determined by $\Pi_{\alpha}$.

In fact not all of the added freedom in projective structure is empirically available: as `second clock effects' are never seen, $\alpha$ really should be exact.\footnote{See \citet{AfriatWeyl} and \citet{EPS}. We thank an anonymous referee for reminding us about second clock effects.} We have to make a choice, and will take affine structure to represent inertia; but if (duly restricted) projective structure is preferred, the arithmetic can be adjusted accordingly.\footnote{The four additional degrees of freedom would be subject to the differential restriction $d\alpha=0$; the two-form $d\alpha$ has six independent quantities.}

\subsection{Curvature and low-dimensional idealisations}\label{midwife}
\noindent Before moving on to the underdetermination of inertia by matter we should consider the extent to which curvature interferes with low-dimensional (zero- or one-dimensional) idealisations that have a role here. We have associated inertia with the geodesics of a connection;\footnote{\emph{Cf}.\ \citet[p.~79]{Lusanna}: ``a global vision of the equivalence principle implies that only global non-inertial frames exist in general relativity [\,\dots]." In other words: \emph{since low-dimensional frames are too small to make sense, they have to be global; global frames are too large to be inertial; hence only non-inertial frames can be countenanced in general relativity}.} and a geodesic is a (parametrised) one-dimensional manifold, a world\emph{line} that (if timelike) can be described by an ideally small---essentially zero-dimensional---object with negligible mass and spatial extension. Masses can be large enough to produce observable distortions of space-time---or small enough to distort only unmeasurably: whatever the threshold of instrumental sensitivity, masses falling below the threshold can always be found. And even if the relationships between the worldlines making up the worldtube of an extended object may not be uninteresting---their geodesic deviation will not always vanish---there will always be geodesics whose separation is small enough to bring geodesic deviation under the threshold of measurability.

\citet[pp.~ix-x]{Synge} was
\begin{quote}
never [\,\dots] able to understand th[e] principle [of equivalence]. [\,\dots] Does it mean that the effects of a gravitational field are indistinguishable from the effects of an observer's acceleration? If so, it is false. In Einstein's theory, either there is a gravitational field or there is none, according as the Riemann tensor does not or does vanish. This is an absolute property; it has nothing to do with any observer's world-line.
\end{quote}
It is doubtless right to distinguish between curvature and flatness; but also between mathematical distinguishability and experimental distinguishability.
\begin{quote}
[\,\dots] The Principle of Equivalence performed the essential office of midwife at the birth of general relativity, but, as Einstein remarked, the infant would never have got beyond its long-clothes had it not been for Minkowski's concept. I suggest that the midwife be now buried with appropriate honours and the facts of absolute space-time faced.
\end{quote}
We suggest more tolerance for the midwife, and certainly not burial; for even in a curved region one can always find a cell (`Einstein's elevator') small enough to make tidal effects experimentally negligible throughout.\footnote{\emph{Cf}.\ \citet[p.~80]{Lusanna}: ``Special relativity can be recovered only locally by a freely falling observer in a neighborhood where tidal effects are negligible," and p.~91: ``[the equivalence principle] suggested [\,\dots] the impossibility to distinguish a uniform gravitational field from the effects of a constant acceleration by means of local experiments in sufficiently small regions where the effects of tidal forces are negligible."} Of course an elevator that's small enough for one level of instrumental sensitivity may not be for another. The strategy is familiar from analysis: for any tolerance $\varepsilon>0$ one can always find a $\delta$ that gives rise to effective indistinguishability by falling under the tolerance. Mathematical physics is full of linear approximations; one often takes the first term in a Taylor expansion and ignores the others.

Tidal effects already get `idealised away' in the sixth corollary (to the laws), where Newton points out that a system of bodies\footnote{``corpora moveantur quomodocunque inter se"} will be indifferent\footnote{``pergent omnia eodem modo moveri inter se, ac si viribus illis non essent incitata," ``corpora omnia æqualiter (quoad velocitatem) movebunt per legem II.\ ideoque nunquam mutabunt positiones et motus eorum inter se."} to a common ``accelerative force."\footnote{``a viribus acceleratricibus æqualibus"} He presumably means a `universal' force subjecting all of them to the same acceleration, and clearly has gravity in mind---which he doesn't mention explicitly, however, as it would produce tidal effects at odds with the claimed invariance. He idealises the difficulty away by specifying conditions that would (strictly speaking) be incompatible if the accelerations were indeed gravitational: they have to be ``equal"\footnote{``æqualibus," ``æqualiter"}---which would put the bodies at the same distance from the source---and in the same direction\footnote{``secundum lineas parallelas"}---which would align them along the same ray. Together the two conditions would confine the bodies to the same spot. Here too, then, there is a sense in which gravity can only be transformed away at a point. The absence of curvature nonetheless makes inertia easier to represent in Newtonian mechanics, where it can be `global' (rather than low-dimensional), since geodesic deviation vanishes everywhere; but as we are wondering to what extent the `relative' inertia of general relativity represents a satisfactory response to the absolute inertia of Newtonian mechanics, we have to represent inertia in general relativity too. Affine structure seems to capture it well---even if real objects are extended and distort space-time.

Then there is the \emph{Wegtransformierbarkeit} of gravitational energy. Though punctual (zero-dimensional) \emph{Wegtransformierbarkeit} has the merit of being logically clean---\emph{some objects satisfy it, others don't}---it may perhaps be too easily satisfied to be meaningful. Larger domains tend to make it harder; they complicate the logic and mathematics of \emph{Wegtransformierbarkeit} by introducing differential constraints tying the fates of certain points to those of others. Curvature might appear to prevent broader \emph{Wegtransformierbarkeit} quite generally, but non-vanishing connection components do not keep $t^{\mu}_{\nu}$ from vanishing: \citet{Schroedinger} proposed coordinates that make $t^{\mu}_{\nu}$ vanish everywhere in an entirely curved universe; so one should not even think of a `bump in the carpet' that can be moved around but not altogether eliminated. As we shall see in \S\ref{reply}, \citet{EinsteinSchroedinger} made Schrödinger's example look pathological by showing that two objects (kept apart by a rod!) are enough to prevent $t^{\mu}_{\nu}$ from vanishing everywhere. But since useful \emph{general} statements (like a satisfactory classification of cases) about how $t^{\mu}_{\nu}$ is affected by coordinate transformations over an arbitrary region seem hard to make, one is tempted to stick to a single point---where the logic of \emph{Wegtransformierbarkeit} is simplified by depending on the object in question alone. Though many quasi-local characterisations of matter-energy have been proposed, they all appear to have their shortcomings; \citet[p.~9]{Szabados} writes:
\begin{quote}
However, contrary to the high expectations of the eighties, finding an appropriate quasi-local notion of energy-momentum has proven to be surprisingly difficult. Nowadays, the state of the art is typically postmodern: Although there are several promising and useful suggestions, we have not only no ultimate, generally accepted expression for the energy-momentum and especially for the angular momentum, but there is no consensus in the relativity community even on general questions (for example, what should we mean e.g. by energy-momentum: only a general expression containing arbitrary functions, or rather a definite one free of any ambiguities, even of additive constants), or on the list of the criteria of reasonableness of such expressions. The various suggestions are based on different philosophies, approaches and give different results in the same situation. Apparently, the ideas and successes of one construction have only very little influence on other constructions.
\end{quote}
The impressive efforts devoted to such constructions are no doubt due to a sense that the legitimacy energy and its conservation rightly have in the rest of physics must be extended to general relativity, however badly they get complicated or even compromised by curvature and path-dependence. Without attempting a serious evaluation of the fruits such efforts have yielded we will confine ourselves to punctual \emph{Wegtransformierbarkeit}, which is mathematically more straightforward and tractable, and logically much cleaner than broader kinds.

The physical significance of tensors is, incidentally, not unrelated to these matters---a tensor being an object that \emph{cannot be transformed away}; but \emph{at a point}. A field that's \emph{wegtransformierbar} at a point may not be more broadly.

\subsection{The underdetermination of inertia by matter}\label{underdetermination}
\noindent We can now try to characterise and quantify the underdetermination, at a point, of inertia by matter. The relationship between affine structure and curvature is given by
$$B^{\mu}_{\nu\kappa\lambda}=2\mathit{\Gamma}^{\mu}_{\nu\textrm{[}\lambda,\kappa\textrm{]}}
+\mathit{\Gamma}^{\tau}_{\nu\lambda}\mathit{\Gamma}^{\mu}_{\tau\kappa}-\mathit{\Gamma}^{\tau}_{\nu\kappa}\mathit{\Gamma}^{\mu}_{\tau\lambda}.$$
The curvature tensor $B^a_{bcd}$ has ninety-six ($6\times 4^2$) independent quantities, eighty if the connection is symmetric, only twenty if it is metric, in which case $B^a_{bcd}$ becomes the Riemann tensor $R^a_{bcd}$. Einstein's equation expresses the equality of the matter tensor $T_{ab}$ and Einstein tensor
$$G_{ab}=R_{ab}-\frac{1}{2}Rg_{ab}\textrm{,}$$
where the Ricci scalar $R$ is the contraction $g^{ab}R_{ab}$ of the Ricci tensor $R_{ab}=R^c_{acb}$. Many Riemann tensors therefore correspond to the same Ricci tensor, to the same Einstein tensor. By removing the ten freedom degrees of a symmetric index pair, the contraction $R_{ab}=R^c_{acb}$ leaves the ten independent quantities of the Ricci tensor; the lost freedoms end up in the Weyl tensor
$$C_{abcd}=R_{abcd}-g_{a\textrm{[}c}R_{d\textrm{]}b}+g_{b\textrm{[}c}R_{d\textrm{]}a}+\frac{1}{3}Rg_{a\textrm{[}c}g_{d\textrm{]}b}.$$
To the disappointment of the relationist, local matter would therefore seem to underdetermine inertia by ten degrees of freedom---some of which may prove less meaningful than others, however, as we shall soon see. But whatever the meaning of the laxity between inertia and matter, their relationship already looks more balanced than before, for now there is \emph{inter}action: besides guiding matter, inertial (\emph{i.e}.\ affine) structure is also constrained by it. Of course this impression of apparent balance or justice, however encouraging, does not settle the issue---the guidance after all leaves no freedom, perhaps the constraint shouldn't either. In fact we still have every reason to wonder about the way matter constrains inertia in general relativity.

Before we see how inertia is constrained by the simplest configuration of matter---its complete absence---in the linear approximation, let us consider a point raised by Ehlers\footnote{\citet[p.~467]{Ehlers}: ``So far, any description of the properties and states of matter involves a metric as an indispensible ingredient. Consequently, quite apart from mathematical technicalities the idea that ``matter determines the metric" cannot even be meaningfully formulated. Besides matter variables, a metric [\,\dots] seems to be needed as an independent, primitive concept in physics [\,\dots]." We thank an anonymous referee for having brought this up.} and others: matter-energy would appear to make no sense without the metric. How can matter-energy underdetermine a more fundamental object that it requires and presupposes?

To begin with, no metric is needed to make sense of one conceptually important matter-energy tensor, namely $T^{ab}=0$. The next-simplest matter-energy tensor is $T^{ab}=\rho V^aV^b$ (`dust'), with matrix representation
$$
\left(
  \begin{array}{cccc}
    \rho & 0 & 0 & 0 \\
    0 & 0 & 0 & 0 \\
    0 & 0 & 0 & 0 \\
    0 & 0 & 0 & 0 \\
  \end{array}
\right).
$$
To rule out tachyonic dust one may seem to need the metric, to impose $g_{ab}V^aV^b<0$; but since conformally equivalent metrics $e^{\lambda}g_{ab}$ all agree, in the sense that
$$[g_{ab}V^aV^b<0]\Leftrightarrow [e^{\lambda}g_{ab}V^aV^b<0]$$
for every $\lambda$, conformal structure is enough. The next-simplest matter-energy tensor is
$$T_{ab}=\rho V_aV_b+p(g_{ab}+V_aV_b)\textrm{,}$$
with matrix
$$
\left(
  \begin{array}{cccc}
    \rho & 0 & 0 & 0 \\
    0 & p & 0 & 0 \\
    0 & 0 & p & 0 \\
    0 & 0 & 0 & p \\
  \end{array}
\right).
$$
The number $p$ typically gets identified with \emph{pressure}, which does involve the metric, being defined as force per unit \emph{area}, or distance squared. The metric is also needed to raise and lower indices: to turn $V^a$ into $V_a$ or $g_{ab}$ into $g^{ab}$, or even (by converting $T_{ab}$ into $T^a_b$) to speak of $\rho$ or $p$ as eigenvalues, or of $V^a$ as an eigenvector. Electromagnetism in general relativity also requires the metric, which appears in the second term of the energy-momentum tensor
$$
T_{ab}=\frac{1}{4\pi}\left(F_{ac}F^c_b-\frac{1}{4}g_{ab}F_{de}F^{de}\right)\textrm{,}
$$
and is also needed to relate $F_{ab}$ or $F^{ab}$ to $F^a_b$. But even if we have decided to represent matter with $T^a_b$ \emph{however} it is constituted, the `materiality' of pure electromagnetism is suspect and open to question; it can be viewed as lower-grade stuff than dust, for instance. And it must be remembered that we are interested in the relationship between matter and \emph{inertia}: admittedly inertia is closely related to the metric in standard general relativity (by $\nabla g=0$); but that relationship, which can be seen as contingent, has been relaxed by \citet{Einstein1925}\footnote{The connection and metric were first varied independently by \citet{Einstein1925}, but he, misled by Pauli, wrongly attributed the method to \citet{Palatini}---who had in fact varied the metric connection; see \citet{FFR}.} and others.

Generally, then, the reliance of matter on the metric seems to depend on the \emph{kind} of matter; in particular on how rich, structured and complicated it is. The simplest matter---absent matter---can do without the metric; the more frills it acquires, the more it will need the metric. We shall continue to explore the underdetermination of inertia by matter, which will be altogether absent in \S\ref{waves} and can otherwise be thought of---with a loss of generality that needn't be too troubling---as a pressureless dust.\footnote{A world made of \emph{dust} or \emph{nothing} may seem a trifle arid. In principle it could be enriched by the six freedom degrees of the symmetric tensor $T_{ij}$, whose eigenvalues $p_1$, $p_2$ and $p_3$ would, if different, indicate a curious spatial anisotropy; to avoid which $T_{ij}$ is taken to be degenerate, with eigenvalue $p=p_1=p_2=p_3$, so that only a single quantity gets typically added to the four of dust. Less arid, but barely \ldots}

To understand how gauge choices eliminate eight degrees of freedom let us now turn to gravitational waves\footnote{For a recent and readable account see \citet{Kennefick}.} in the linear approximation.

\subsection{Inertia without matter}\label{waves}
\noindent Through Einstein's equation, then, matter determines the rough curvature given by the Ricci tensor. The \emph{absence} of matter,\footnote{\citet[p.~132]{KosmologischeBetrachtungen}, it is worth mentioning, wrote that without matter there is no inertia at all: ``In a consistent theory of relativity there can be no inertia \emph{with respect to `space\emph{,}'} but only an inertia of the masses \emph{with respect to one another}."} for instance, makes that curvature vanish identically---but not the more detailed Riemann curvature, which can oscillate nonetheless, and in many different ways. Here we will see the purely `Weyl' freedom left by the absence of matter.

The weak perturbation $h_{\mu\nu}=g_{\mu\nu}-\eta_{\mu\nu}$ would (being symmetrical) first appear to maintain the ten freedoms of the Weyl tensor. It is customary to write
$\gamma_{\mu\nu}=h_{\mu\nu}-\frac{1}{2}\eta_{\mu\nu}h\textrm{,}$ where $h$ is the trace $h^{\mu}_{\mu}$. A choice of coordinates satisfying the four continuity conditions $\partial_{\nu}\gamma_{\mu\nu}=0$ allows us to set $\gamma_{\mu 0}=0$, which does away with the four `temporal' freedoms. There remains a symmetric `purely spatial' matrix
$$
\left(
  \begin{array}{cccc}
    0 & 0 & 0 & 0 \\
    0 & \gamma_{11} & \gamma_{21} & \gamma_{31} \\
    0 & \gamma_{21} & \gamma_{22} & \gamma_{32} \\
    0 & \gamma_{31} & \gamma_{32} & \gamma_{33} \\
  \end{array}
\right)
$$
with six degrees of freedom. We can also make $h$ vanish, which brings us back to $h_{\mu\nu}=\gamma_{\mu\nu}$ and eliminates another freedom, leaving five. To follow the fates of these remaining freedoms we can consider the plane harmonic
\begin{equation}\label{harmonic}
h_{\mu\nu}=\textrm{Re}\{A_{\mu\nu}e^{\textrm{i}\langle k,x\rangle}\}
\end{equation}
obeying $\square h_{\mu\nu}=0$. If the wave equation were $\square_\textsf{c}h_{\mu\nu}=(\partial_0^2-\textsf{c}^2\nabla^2)h_{\mu\nu}=0$ instead, with arbitrary $\textsf{c}$, the wave (co)vector $k$ would have four independent components $k_{\mu}~\negthinspace=~\negthinspace\langle k,\partial_{\mu}\rangle$:
\begin{itemize}
\item the direction $k_1:k_2:k_3$, in other words $\mathbf{k}/|\mathbf{k}|$ (two)
\item the length $|\mathbf{k}|=\sqrt{k_1^2+k_2^2+k_3^2}$ (one)
\item the frequency $\omega =k_0=\langle k,\partial_0\rangle=\textsf{c}|\mathbf{k}|$ (one).
\end{itemize}
Since $\textsf{c}=1$ is a natural constant, the condition $\square h_{\mu\nu}=0$ reduces them to three, by identifying $|\mathbf{k}|$ and $\omega$, which makes the squared length $k_ak^a=k_0k^0-|\mathbf{k}|^2$ vanish. And even these three degrees of freedom disappear into the coordinates if the wave is made to propagate along the third spatial axis, which can be recalibrated to match the wavelength, leaving two ($5-3$) freedoms, of polarisation. The three orthogonality relations
$$\sum_{j=1}^3A_{ij}k^j=\sum_{j=1}^3A(\partial_i,\partial_j)\langle dx_a^j,k^a\rangle=0$$
($i=1,2,3$) follow from $\partial_{\nu}\gamma_{\mu\nu}=0$ and situate the polarisation tensor $A$ with components $A_{ij}$ in the plane $\mathbf{k}^{\perp}\subset k^{\perp}$ orthogonal to the three-vector $\mathbf{k}\in k^{\perp}$. Once the coordinates are realigned and recalibrated so that $\langle\mathbf{k},\partial_3\rangle=1$ and $\langle\mathbf{k},\partial_1\rangle$, $\langle\mathbf{k},\partial_2\rangle$ both vanish, the three components $A(\partial_3,\partial_j)$ also vanish, leaving a traceless symmetric matrix
$$\left(
  \begin{array}{cccc}
    0 & 0 & 0 & 0 \\
    0 & h_{11} & h_{21} & 0 \\
    0 & h_{21} & -h_{11} & 0 \\
    0 & 0 & 0 & 0 \\
  \end{array}
\right)
$$
with two independent components, $h_{11}=-h_{22}$ and $h_{12}=h_{21}$.

The above gauge choices therefore eliminate eight degrees of freedom:
\begin{itemize}
\item the four `temporal' coordinates $\gamma_{\mu 0}$ eliminated by the conditions $\partial_\nu\gamma_{\mu\nu}=0$
\item the freedom eliminated by $h=0$
\item the three freedoms of $\mathbf{k}$ eliminated by realignment and recalibration.
\end{itemize}

One may wonder about the use of an only `linearly' covariant approximation in a paper that so insistently associates physical legitimacy with \emph{general} covariance. The linear approximation has been adopted as the simplest way of illustrating how two degrees of freedom remain after gauge choices eliminate eight; but the same count ($2=10-8$) can be shown, though much less easily, to hold in general. Very briefly: The ten vacuum field equations $G_{\mu\nu}=0$ are not independent, being constrained\footnote{See \citet{RyckmanBrading} and \citet{Ryckman2008} on Hilbert's struggle, with similar constraints, to reconcile causality and general covariance.} by the four contracted Bianchi identities $\nabla_aG^{a0}=\cdots=\nabla_aG^{a3}=0$; another four degrees freedom are lost to constraints on the initial data, leaving two.\footnote{In the general nonlinear case the two remaining freedoms can be harder to recognise as polarisations of gravitational waves; Lusanna \& Pauri speak of the ``two autonomous degrees of freedom of the gravitational field." But having based our arithmetic on the linear approximation we will continue to speak of polarisation.} For details we refer the reader to \citet[pp.~95-6]{Lusanna}, \citet[pp.~696, 699, 706-7]{LusannaPauri} and \citet[pp.~193-4]{LusannaPauri2}; but should point out that their (related) agenda makes them favour a different, `canonical' (or `double') arithmetic ($2\cdot2=2\cdot\textrm{[}10-4-4\textrm{]}$) of freedom degrees provided by the ADM Hamiltonian formalism, which they use to distinguish between four---two configurational and two canonically conjugate---``ontic" (or ``tidal" or ``gravitational") quantities and the remaining ``epistemic" (or ``inertial" or ``gauge") degrees of freedom.\footnote{Lusanna \& Pauri also take the four eigenvalues of the Weyl tensor, and gravitational `observables' characterised in various ways by Bergmann \& Komar, to express `genuine gravity' as opposed to mere `inertial appearances.'} The ontic-tidal-gravitational quantities---the \emph{Dirac observables}---are not numerically invariant\footnote{But \citet[p.~101]{Lusanna}: ``Conjecture: there should exist privileged Shanmugadhasan canonical bases of phase space, in which the DO (the tidal effects) are also \emph{Bergmann observables}, namely coordinate-independent (scalar) tidal effects."} under the group $\mathcal{G}_8$ of gauge transformations; Lusanna \& Pauri seem to view a gauge choice $\mathit{\Gamma}_8\in\mathcal{G}_8$ as determining a specific realisation (or `coordinatisation'?) `$\Omega_4=\mathit{\Gamma}_8(\tilde{\Omega}_4)$' of a single ``abstract" four-dimensional symplectic space $\tilde{\Omega}_4$.\footnote{See \citet[pp.~706-7]{LusannaPauri}; and also \citet[p.~101]{Lusanna}: ``The reduced phase space of this model of general relativity is the space of abstract DO (pure tidal effects without inertial effects), which can be thought as four fields on an abstract space-time $\tilde{M}^4=$ {\emph{equivalence class of all the admissible non-inertial frames $M^4_{3+1}$ containing the associated inertial effects}}."} The ontic state can perhaps be understood as an invariant point $\omega\in\tilde{\Omega}_4$, which acquires the four components $\{q^1\textrm{(}\omega\textrm{)},\dots,p_2\textrm{(}\omega\textrm{)}\}\in\Omega_4$ with respect to the coordinates $q^1,q^2,p_1,p_2$ characterising a particular $\Omega_4$. At any rate, Lusanna \& Pauri use the \emph{four ontic-tidal-gravitational observables} to
\begin{itemize}
\item individuate space-time points
\item `dis-solve' the hole argument\footnote{They point out that the diffeomorphism at issue is constrained by the fixed Cauchy data to be purely `epistemic' and not `ontic'; the covariance is not \emph{general}.}
\item argue that change \emph{is} possible in canonical gravity, for the `ontic' quantities \emph{can} evolve.\footnote{The Hamiltonian acting on the reduced phase space is not constant in asymptotically flat space-times, where consistency requires the addition of a (De Witt surface) term generating a genuine `ontic' evolution; see \citet[p.~97]{Lusanna}, for instance. \emph{Cf}.\ \citet[\S\S 4-6]{BelotEarman2001} for a complementary discussion of time and change in canonical gravity; or \citet{Earman2006} p.~451: ``In the case of GTR the price of saving determinism is a frozen picture of the world in which the observables do not change over time." Where the space-time is spatially compact, with no boundary, the `ontic' quantities---Earman speaks of ``observables"---remain unchanged as no surface term has to be added to the (constant `ontic') Hamiltonian governing their evolution. Time evolution $X_H=(dH)^\#$ is after all generated by the \emph{differential} $dH$ of the Hamiltonian, which vanishes if the Hamiltonian is constant---for instance if it vanishes identically. We thank an anonymous referee for added precision on this matter.}
\end{itemize}
Since so much hangs on their four observables, Lusanna \& Pauri emphasise---with detailed metrological considerations---that \emph{they really are observable}, and go into possible schemes for their measurement. In \S\ref{Doppler} we propose a Doppler effect in a similar spirit; but let us now return to the two (configurational, as opposed to canonical) degrees of freedom left by the eight gauge choices.

The physical meaning of coordinate transformations has been amply discussed, notably in the literature on the hole argument.\footnote{See \citet{EarmanNorton}, \citet{Butterfield87,Butterfield}, \citet{Norton}, \citet[\S9]{Earman}, \citet{Maudlin}, \citet{Stachel}, \citet{Rynasiewicz1}, \citet{Belot1996}, \citet{BelotEarman1999,BelotEarman2001}, \citet[pp.\ 19-23]{RyckmanReign}, \citet{Earman2006}, \citet{DoratoPauri}, \citet{LusannaPauri}, \citet[pp.~99-100]{Lusanna}, \citet[\S2.2.5]{Rovelli}, \citet[\S2]{EsfeldLam} for instance.} The relationist will take the eight degrees freedom eliminated by the above gauge choices to be meaningless,\footnote{\emph{Cf}.\ \citet[\S2.3.2]{Rovelli}.} to lessen the underdetermination of inertia---and because as a relationist he would anyway. We will too, and concentrate on the status of the double freedom of polarisation.\footnote{\emph{Cf}.\ \citet{Earman2006} p.~444: ``In what could be termed the classical phase of the debate, the focus was on coordinate systems and the issue of whether equations of motion/field equations transform in a generally covariant manner under an arbitrary coordinate transformation. But from the perspective of the new ground the substantive requirement of general covariance is not about the status of coordinate systems or covariance properties of equations under coordinate transformation; indeed, from the new perspective, such matters cannot hold any real interest for physics since the content of space-time theories [\,\dots] can be characterised in a manner that does not use or mention coordinate systems. Rather, the substantive requirement of general covariance lies in the demand that diffeomorphism invariance is a gauge symmetry of the theory at issue." A distinction between \emph{physically meaningful} and \emph{mere gauge} is at the heart of the new perspective. \emph{Cf}.\ \citet[p.~104]{Lusanna}: ``the true physical degrees of freedom [\,\dots] are the gauge invariant quantities, the \emph{Dirac observables} (DO)."}

Matter still underdetermines inertia, then, by two degrees freedom, which obstruct the satisfaction of `Mach's principle,' as we are calling it. But are they really there? Or do they share the fate of the eight freedoms eliminated by gauge choices, which we have dismissed as physically meaningless? The relationist may prefer to discard them too as an empty mathematical fiction without physical consequence; but we know their physical meaning is bound up with that of gravitational waves, whose polarisation they represent.

We should emphasise that the formalism of general relativity (especially in its Lagrangian and Hamiltonian versions) distinguishes clearly between the eight degrees of freedom eliminated by gauge choices and the remaining two representing polarisation. We are not claiming that all ten ($=8+2$) are theoretically, mathematically on an equal footing, for they are not; we are merely wondering about the physical meaning of the two that cannot be eliminated by gauge choices. The claim that the physical meaning of these two polarisations is related to the status of gravitational waves seems relatively uncontentious---unlike the much more alarming claim (which we are undeniably assessing, but not making) that gravitational waves are a physically unreal mathematical fiction.

\subsection{Gravitational waves, transformation behaviour and reality}\label{R&I}

\noindent To deal with the polarisation obstructing a full determination of inertia the relationist can insist on the right transformation behaviour, which gravitational waves do not seem to have, in various senses. He will argue that as the generation and energy, perhaps even the detection of gravitational waves can be transformed away, they and the underdetermination of inertia by matter are about as fictitious as the eight freedoms that have just disappeared into the coordinates.

If gravitational waves had mass-energy their reality could be hard to contest.\footnote{\emph{Cf}.\ \citet{Dorato2000}: ``Furthermore, the gravitational field has momentum energy, therefore mass (via the equivalence between mass and energy) and having mass is a typical feature of substances."} We have seen that general relativity does allow the attribution of mass-energy to the gravitational field, to gravitational waves, through the pseudotensor $t^{\mu}_{\nu}$; but also that $t^{\mu}_{\nu}$ has the wrong transformation behaviour.

Is the physical meaning of $t^{\mu}_{\nu}$ really compromised by its troubling susceptibility\footnote{This issue is logically straightforward at a single point, where it only depends on the object in question (here $t^{\mu}_{\nu}$); the logic of broader \emph{Wegtransformierbarkeit} is much messier, depending on the nature of the region, the presence of cosmic rods \emph{etc}.; see \S\ref{midwife}, and Einstein's reply in \S\ref{reply}.} to disappear, and reappear under acceleration? A similar question arose in \S\ref{matter}, when we wondered what to count as matter. There we did not provide the relationist with the `gravitational matter' that would have favoured his agenda by making his principles easier to satisfy, on grounds that, being mere `opinion,' it was too insubstantial and tenuous to count. To be fair to the relationist we should perhaps dismiss $t^{\mu}_{\nu}$ once more as mere opinion. But we have no reason to be fair, and are merely exploring certain logical possibilities. Perhaps `matter' was something stronger, and required more; maybe a quantity that comes and goes with the accelerations of the observer can be real despite being immaterial; so we shall treat the physical meaning of $t^{\mu}_{\nu}$---as opposed to its suitability for the representation of matter---as a further issue.

General relativity has been at the centre of a tradition, conspicuously associated with \citet[pp.~261 (Teil I), 276-8 (Teil II)]{Hilbert},\footnote{See also \citet{RyckmanBrading} and \citet{Ryckman2008}.} \citet[p.~382]{Levi-Civita}, \citet[pp.~6-7; 1926, p.~492]{Schroedinger}, \citet{Cassirer}, \citet[pp.~5,~13]{Einstein1922} himself eventually, \citet[pp.\ 31, 54]{Langevin}, \citet[\S48]{DedRel}, \citet[\S VII]{Russell1927} and \citet[\S17]{PdMuN}, linking physical reality or objectivity or significance to appropriate transformation properties, to something along the lines of invariance or covariance.\footnote{Covariance and invariance are rightly conflated in much of the literature, and here too. Whether it is a number or \emph{Gestalt} or syntax or the appearance of a law that remains unchanged is less the point than the generality---\emph{complete} or \emph{linear} or \emph{Lorentz}, for instance---of the transformations at issue.} Roots can be sought as far back as Democritus, who is said to have claimed that ``sweet, bitter, hot, cold, colour" are mere opinion, ``only atoms and void"---concerning which there ought in principle to be better agreement---``are real"; or more recently in Felix Klein's `Erlangen programme' \citeyearpar{Klein}, which based geometrical relevance on invariance under the groups he used to classify geometries. Bertrand Russell, in his version of neutral monism,\footnote{Accounts can be found in Russell (\citeyear{Russell1921}, \citeyear{Russell1927}, \citeyear{Russell1956}). But see also \citet[p.\ 14]{Russell1912}, which was first published in 1912. \emph{Cf}.\ \citet[p.\ 36]{Cassirer}.} identified objects with the class of their appearances from different points of view---not really an association of invariance and reality, but an attempt to transcend the misleading peculiarities of individual perspectives nonetheless. Hilbert explicitly required invariance in ``Die Grundlagen der Physik," denying physical significance to objects with the wrong transformation properties. Levi-Civita, \citet{Schroedinger} and \citet[p.~165]{Bauer}, who saw the relation of physical meaning to appropriate transformation properties as a central feature of relativity theory, likewise questioned\footnote{See \citet{CattaniDeMaria}.} the significance of the energy-momentum pseudotensor. Schrödinger noted that appropriate coordinates make $t^{\mu}_{\nu}$ vanish identically in a curved space-time (containing only one body); Bauer that certain `accelerated' coordinates would give energy-momentum to flat regions.

Einstein first seemed happy to extend physical meaning to objects with the wrong transformation properties. In January 1918 he upheld the reality of $t^{\mu}_{\nu}$ in a paper on gravitational waves:
\begin{quote}
[Levi-Civita] (and with him other colleagues) is opposed to the emphasis of equation [$\partial_{\nu}(\mathfrak{T}^{\nu}_{\sigma}+\mathfrak{t}^{\nu}_{\sigma})=0$] and against the aforementioned interpretation, because the $\mathfrak{t}^{\nu}_{\sigma}$ do not make up a \emph{tensor}. Admittedly they do not; but I cannot see why physical meaning should only be ascribed to quantities with the transformation properties of tensor components.\footnote{\citet[p.\ 167]{EinsteinGravitationswellen}: ``[Levi-Civita] (und mit ihm auch andere Fachgenossen) ist gegen eine Betonung der Gleichung [$\partial_{\nu}(\mathfrak{T}^{\nu}_{\sigma}+\mathfrak{t}^{\nu}_{\sigma})=0$] und gegen die obige Interpretation, weil die $\mathfrak{t}^{\nu}_{\sigma}$ keinen \so{Tensor} bilden. Letzteres ist zuzugeben; aber ich sehe nicht ein, warum nur solchen Größen eine physikalische Bedeutung zugeschrieben werden soll, welche die Transformationseigenschaften von Tensorkomponenten haben."\label{Gravitationswellen}}
\end{quote}
\subsection{Einstein's reply to Schrödinger}\label{reply}
\noindent In February \citeyearpar{EinsteinSchroedinger} Einstein responded to Schrödinger's objection, arguing that with more than one body the stresses $t^i_j$ transmitting gravitational interactions would not vanish: Take two bodies $M_1$ and $M_2$ kept apart by a rigid rod $R$ aligned along $\partial_1$. $M_1$ is enclosed in a two-surface $\partial\Theta$ which leaves out $M_2$ and hence cuts $R$ (orthogonally one can add, for simplicity). Integrating over the three-dimensional region $\Theta$, the conservation law $\partial_{\nu} U^{\nu}_{\mu}=0$ yields
$$\frac{d}{dx^0}\int_{\Theta}U^0_{\mu}d^3\mathbf{x}=\int_{\partial\Theta}\sum_{i=1}^3U^i_{\mu}d^2\Sigma_i:$$
 any change in the total energy $\int U^0_{\mu}d^3\mathbf{x}$ enclosed in $\Theta$ would be due to a flow, represented on the right-hand side, through the boundary $\partial\Theta$ (where $U^{\mu}_{\nu}$ is again $T^{\mu}_{\nu}+t^{\mu}_{\nu}$, and $d^3\mathbf{x}$ stands for $dx^1\wedge dx^2\wedge dx^3$; we have replaced Einstein's cosines with a notation similar to the one used, for instance, in \citet{MTW}). Since the situation is stationary and there are no flows, both sides of the equation vanish, for $\mu =0,1,2,3$. Einstein takes $\mu =1$ and uses
 $$\int_{\partial\Theta}\sum_{i=1}^3U^i_1 d^2\Sigma_i=0.$$
He is very concise, and leaves out much more than he writes, but we are presumably to consider the intersection $R\cap\partial\Theta$ of rod and enclosing surface, where it seems $\partial_1$ is orthogonal to $\partial_2$ and $\partial_3$, which means the off-diagonal components $T^2_1$ and $T^3_1$ vanish, unlike the component $T^1_1$ along $R$. Since
$$-\int_{\partial\Theta}\sum_{i=1}^3t^i_1 d^2\Sigma_i$$
must be something like $T^1_1$ times the sectional area of $R$, the three gravitational stresses $t^i_1$ cannot all vanish identically. The argument is swift, contrived and full of gaps, but the conclusion that gravitational stresses between two (or more) bodies cannot be `transformed away' seems valid.

Then in May we again find Einstein lamenting that
\begin{quote}
Colleagues are opposed to this formulation [of conservation] because ($\mathfrak{U}^{\nu}_{\sigma}$) and ($\mathfrak{t}^{\nu}_{\sigma}$) are not tensors, while  they expect all physically significant quantities to be expressed by scalars or tensor components.\footnote{\citet[p.\ 447]{EinsteinEnergiesatz}: ``Diese Formulierung stößt bei den Fachgenossen deshalb auf Widerstand, weil ($\mathfrak{U}^{\nu}_{\sigma}$) und ($\mathfrak{t}^{\nu}_{\sigma}$) keine Tensoren sind, während sie erwarten, daß alle für die Physik bedeutsamen Größen sich als Skalare und Tensorkomponenten auffassen lassen müssen."\label{Energie}}
\end{quote}
In the same paper he defends his controversial energy conservation law,\footnote{See \citet{Hoefer2} on the difficulties of energy conservation.} which we shall soon come to.

\subsection{Conservation under coordinate substitutions}\label{conservation}
\noindent Conservation is bound to cause trouble in general relativity. The idea usually is that even if the conserved quantity---say a `fluid' with density $\rho$---doesn't stay put, even if it moves and gets transformed, an appropriate total over space nonetheless persists through time; a spatial integral remains constant:
\begin{equation}\label{integrallaw}
\frac{d}{dt}\int\negthinspace\rho\thinspace d^3\mathbf{x}=0.
\end{equation}
So a \emph{clean} separation into \emph{space} (across which the integral is taken) and \emph{time} (in the course of which the integral remains unchanged) seems to be presupposed when one speaks of conservation. In relativity the separation suggests a Minkowskian orthogonality
\begin{equation}\label{orthogonality}
\partial_0\perp\textrm{span}\{\partial_1,\partial_2,\partial_3\}
\end{equation}
between time and space,\footnote{\emph{Cf}.\ \citet[p.\ 450]{EinsteinEnergiesatz}.} which already restricts the class of admissible transformations and hence the generality of any covariance. However restricted, the class will be far from empty; and what if the various possible integrals it admits give different results? Or if some are conserved and others aren't?

An integral law like (\ref{integrallaw}) can typically be reformulated as a `local' divergence law
$$\frac{\partial\rho}{\partial t}+\mathbf{\nabla}\cdot\mathbf{j}=0\textrm{,}$$
which in four dimensions reads $\partial_{\mu}J^{\mu}=0$, where $\mathbf{j}$ stands for the current density $\rho\mathbf{v}$, the three-vector $\mathbf{v}$ represents the three-velocity of the fluid, $J^0$ is the density $\rho$ and $J^i$ equals $\langle dx^i,\mathbf{j}\rangle$. But the integral law is \emph{primary}; the divergence law \emph{derived from it} only really expresses conservation to the extent that it is fully equivalent to the more fundamental integral law. As Einstein puts it:
\begin{quote}
From the physical point of view this equation [${\partial\mathfrak{T}^{\nu}_{\sigma}}/{\partial x_{\nu}}+\frac{1}{2}g^{\mu\nu}_{\sigma}\mathfrak{T}_{\mu\nu}=0$] cannot be considered completely equivalent to the conservation laws of momentum and energy, since it does not correspond to integral equations which can be interpreted as conservation laws of momentum and energy.\footnote{\citet[p.\ 449]{EinsteinEnergiesatz}: ``Vom physikalischen Standpunkt aus kann diese Gleichung nicht als vollwertiges Äquivalent für die Erhaltungssätze des Impulses und der Energie angesehen werden, weil ihr nicht Integralgleichungen entsprechen, die als Erhaltungssätze des Impulses und der Energie gedeutet werden können."}
\end{quote}
In flat space-time, with inertial coordinates, the divergence law $\partial_{\mu}T^{\mu}_{\nu}=0$ can be unambiguously integrated to express a legitimate conservation law. But the ordinary divergence $\partial_{\mu}T^{\mu}_{\nu}$ only vanishes in free fall (where it coincides with $\nabla_a T^a_{\nu}$), and otherwise registers the gain or loss seen by an accelerated observer. If such variations are to be viewed as exchanges with the environment and not as definitive acquisitions or losses, account of them can be taken with $t^{\mu}_{\nu}$, which makes $\partial_{\mu}(T^{\mu}_{\nu}+t^{\mu}_{\nu})$ vanish by compensating the difference.\footnote{\emph{Cf}.\ \citet[p.\ 136]{RyckmanBrading}.} The generally covariant condition $\partial_{\mu}(T^{\mu}_{\nu}+t^{\mu}_{\nu})=0$, which is equivalent to $\nabla_a T^a_{\nu}=0$ and $\partial_{\mu}T^{\mu}_{\nu}+\frac{1}{2}\partial_{\nu}g^{ab}T_{ab}=0$, can also be unambiguously integrated in flat space-time to express a legitimate conservation law. But integration is less straightforward in curved space-time, where it involves a distant comparison of direction which cannot be both generally covariant and integrable.

Nothing prevents us from comparing the values of a genuine scalar at distant points. But we know the density of mass-energy transforms according to
$$\textrm{(}\rho,\mathbf{0}\textrm{)}\mapsto\frac{\rho}{\sqrt{1-|\mathbf{v}|^2}} \textrm{(}1,\mathbf{v}\textrm{)}\textrm{,}$$
where $\mathbf{v}$ is the three-velocity of the observer. So the invariant quantity is not the mass-energy density, but (leaving aside the stresses that only make matters worse) the mass-energy-momentum density, which \emph{is manifestly directional}. And how are distant directions to be compared? Comparison of components is not invariant: directions or rather component ratios equal with respect to one coordinate system may differ in another. Comparison by parallel transport will depend not on the coordinate system, but on the path followed.

\subsection{Einstein's defence of energy conservation}\label{defence}
\noindent Einstein tries to get around the problem in ``Der Energiesatz in der allgemeinen Relativitätstheorie" \citeyearpar{EinsteinEnergiesatz}. Knowing that conservation is unproblematic in flat space-time, where parallel transport is integrable, he makes the universe look as Minkowskian as possible by keeping all the mass-energy spoiling the flatness neatly circumscribed (which is already questionable, for matter may be infinite).

Einstein attributes an energy-momentum $J$ to the universe, which he legitimates by imposing a kind of `general' (but in fact restricted) invariance on each component $J_{\mu}$, defined as the spatial integral
$$J_{\mu}=\int\mathfrak{U}^0_{\mu}d^3\mathbf{x}$$ of the combined energy-momentum $\mathfrak{U}^0_{\mu}=\mathfrak{T}^0_{\mu}+\mathfrak{t}^0_{\mu}$ of matter and field (where $\mathfrak{U}^{\mu}_{\nu}=U^{\mu}_{\nu}\sqrt{-\mathbf{g}}$ \emph{etc}., and the stresses seem to be neglected). To impose it he separates time and space through (\ref{orthogonality}), and requires the fields $\mathfrak{T}^{\mu}_{\nu}$ and $\mathfrak{t}^{\mu}_{\nu}$ to vanish outside a bounded region $B$. Einstein is prudently vague about $B$, which is first a subset of a simultaneity slice $\mathit{\Sigma}_t$, and then gets ``infinitely extended in the time direction,"\footnote{\citet[p.\ 450]{EinsteinEnergiesatz}} to produce the world tube $B_{\partial_0}$ described by $B$ along the integral curves of the ``time direction" $\partial_0.$ The supports $\overline{\mathfrak{T}}$ and $\overline{\mathfrak{t}}$ of $\mathfrak{T}^{\mu}_{\nu}$ and $\mathfrak{t}^{\mu}_{\nu}$ are contained in $B_{\partial_0}$ by definition; but $\overline{\mathfrak{T}}$ may be much smaller than $\overline{\mathfrak{t}}$ and hence $B_{\partial_0}$: we have no reason to assume that $\overline{\mathfrak{T}}$ does not contain bodies that radiate gravitational waves---of which $\mathfrak{t}^{\mu}_{\nu}$ would have to take account---along the lightcones delimiting the causal future of $\overline{\mathfrak{T}}_t=\overline{\mathfrak{T}}\cap\mathit{\Sigma}_t$. Gravitational waves could therefore, by obliging $B_{\partial_0}$ to be much larger than $\overline{\mathfrak{T}}$, spoil the picture of an essentially Minkowskian universe barely perturbed by the `little clump' of matter-energy it contains.

The generality of any invariance or covariance is already limited by (\ref{orthogonality}); Einstein restricts it further by demanding Minkowskian coordinates $g_{\mu\nu}=\eta_{\mu\nu}$ (and hence flatness) outside $B_{\partial_0}$.\footnote{Flatness cannot reasonably be demanded of the rest of the universe, as can be seen by giving $T^a_b$ the spherical support it has in the Schwarzschild solution, where curvature diminishes radially without ever vanishing.} He then uses the temporal constancy ${dJ_{\mu}}/{dx^0}=0$ of each component $J_{\mu}$, which follows from $\partial_{\mu}\mathfrak{U}^{\mu}_{\nu}=0$, to prove that $J_{\mu}$ has the same value $(J_{\mu})_1=(J_{\mu})_2$ on both three-dimensional simultaneity slices\footnote{For a recent treatment see \citet{MLR}.} $x^0=t_1$ and $x^0=t_2$ of coordinate system $K$; and value $(J'_{\mu})_1=(J'_{\mu})_2$ at $x^{\prime 0}=t'_1$ and $x^{\prime 0}=t'_2$ in another system $K'$. A third system $K''$ coinciding with $K$ around the slice $x^0=t_1$ and with $K'$ around $x^{\prime 0}=t'_2$ allows the comparison of $K$ and $K'$ across time. The invariance of each component $J_{\mu}$ follows from $(J_{\mu})_1=(J'_{\mu})_2$. Having established that, Einstein views the world as a `body' immersed in an otherwise flat space-time, whose energy-momentum $J_{\mu}$ is covariant under the transformation laws---Lorentz transformations---considered appropriate\footnote{Despite \citet{Kretschmann}, who pointed out that even an entirely flat universe can be considered subject to general (and not just Lorentz) covariance. \emph{Cf}.\ \citet[\S2.4.3]{Rovelli}.} for that (largely flat) environment. Unusal mixture of transformation properties: four components, each one `somewhat' invariant, which together make up a four-vector whose Lorentz covariance would be of questionable appropriateness even if the universe were \emph{completely} flat.

Einstein's argument was nonetheless effective, and persuaded\footnote{See \citet{CattaniDeMaria}, \citet{Hoefer2}.} the community, which became and largely remains more tolerant of objects (including laws and calculations) with dubious transformation properties.

In \S\S\ref{R&I}-8 we saw what Einstein thought in the first months of 1918. Already in ``Dialog über Einwände gegen die Relativitätstheorie," which came out in November, there's a shift, a timid concession to his opponents, a subtler tolerance. Einstein gives the impression\footnote{\citet{Dialog}, middle of second column} he may have been glad to do away with coordinates, if possible---but like Cassirer\footnote{\citet[p.~37]{Cassirer}} he thought it wasn't: ``[\,\dots] cannot do without the coordinate system [\,\dots]."\footnote{\citet[p.~699]{Dialog}: ``Die wissenschaftliche Entwicklung aber hat diese Vermutung nicht bestätigt. Sie kann das Koordinatensystem nicht entbehren, muß also in den Koordinaten Größen verwenden, die sich nicht als Ergebnisse von definierbaren Messungen auffassen lassen."} If he had known\footnote{Bertrand \citet[p.~71]{Russell1927} was perhaps the first to see the possibility of a formulation we would now call `intrinsic' or `geometrical': ``Reverting now to the method of tensors and its possible eventual simplification, it seems probable that we have an example of a general tendency to over-emphasise numbers, which has existed in mathematics ever since the time of Pythagoras, though it was temporarily less prominent in later Greek geometry as exemplified in Euclid. [\,\dots] Owing to the fact that arithmetic is easy, Greek methods in geometry have been in the background since Descartes, and co-ordinates have come to seem indispensable. But mathematical logic has shown that number is logically irrelevant in many problems where it formerly seemed essential [\,\dots]. A new technique, which seems difficult because it is unfamiliar, is required when numbers are not used; but there is a compensating gain in logical purity. It should be possible to apply a similar process of purification to physics. The method of tensors first assigns co-ordinates, and then shows how to obtain results which, though expressed in terms of co-ordinates, do not really depend upon them. There must be a less indirect technique possible, in which we use no more apparatus than is logically necessary, and have a language which will only express such facts as are now expressed in the language of tensors, not such as depend on the choice of co-ordinates. I do not say that such a method, if discovered, would be preferable in practice, but I do say that it would give a better expression of the essential relations, and greatly facilitate the task of the philosopher."} that one can write, say, $\nabla V$ instead of
\begin{equation}\label{covariant}
\partial_{\mu}V^{\nu}+\mathit{\Gamma}^{\nu}_{\mu\kappa}V^{\kappa}\textrm{,}
\end{equation}
Einstein would simply have attributed `full' reality to $\nabla V$ (without bothering with confusing compromises). But he saw the complicated compensation of expressions like (\ref{covariant}) instead, in which various transformations balance each other to produce a less obvious invariance: ``Only certain, generally rather complicated expressions, made up of field components and coordinates, correspond to coordinate-independent measurable (\emph{i.e.} real) quantities."\footnote{\citet[p.~699-700]{Dialog}: ``Nur gewissen, im allgemeinen ziemlich komplizierten Ausdrücken, die aus Feldkomponenten und Koordinaten gebildet werden, entsprechen vom Koordinatensystem unabhängig meßbare (d.~h.\ reale) Größen." A similar idea is expressed in \citet[p.~278, \so{Drittens}. \dots]{Hilbert}; \emph{cf}.\ \citet[p.~136]{RyckmanBrading}: ``Interestingly, Hilbert here cites the example of energy in general where the (`pseudo-tensor density') expression for the energy-momentum-stress of the gravitational field is not generally invariant but nonetheless, if defined properly, occurs in the statement of a
conservation law that holds in every frame, i.e., is generally covariant."} He felt that ``the gravitational field [$\mathit{\Gamma}^{\mu}_{\nu\kappa}$] at a point is neither real nor merely fictitious"\footnote{\citet[p.~700]{Dialog}: ``\foreignlanguage{german}{Man kann deshalb weder sagen, das Gravitationsfeld an einer Stelle sei etwas "`Reales"', noch es sei etwas "`bloß Fiktives"'.}"}: not entirely real since it has ``part of the arbitrariness"\footnote{\emph{Ibid}.\ p.~699: ``Nach der allgemeinen Relativitätstheorie sind die vier Koordinaten des raum-zeitlichen Kontinuums sogar ganz willkürlich wählbare, jeder selbständigen physikalischen Bedeutung ermangelnde Parameter. Ein Teil jener Willkür haftet aber auch denjenigen Größen (Feldkomponenten) an, mit deren Hilfe wir die physikalische Realität beschreiben."} of coordinates; not fictitious because it participates---``the field components [\,\dots] with whose help we describe physical reality"---in the balancing act yielding invariant reality: ``nothing `physically real' corresponds to the gravitational field \emph{at a point}, only to the gravitational field in conjunction with other data."\footnote{\emph{Ibid}.\ p.~700: ``\foreignlanguage{german}{dem Gravitationsfeld \emph{an einer Stelle} entspricht also noch nichts "`physikalisch Reales"', wohl aber diesem Gravitationsfelde in Verbindung mit anderen Daten.}"}

\subsection{Einstein's conversion}\label{conversion}
\noindent In May 1921 Einstein seems to have gone a good deal farther, approaching, perhaps even exceeding the positions of his former opponents:
\begin{quote}With the help of speech, different people can compare their experiences to a certain extent. It turns out that some---but not all---of the sensory experiences of different people will coincide. To such sensory experiences of different people which, by coinciding, are superpersonal in a certain sense, there corresponds a reality. The natural sciences, and in particular the most elementary one, physics, deal with that reality, and hence indirectly with the totality of such experiences. To such relatively constant experience complexes corresponds the concept of the physical body, in particular that of the rigid body.\footnote{\citet[p.\ 5]{Einstein1922}: ``Verschiedene Menschen können mit Hilfe der Sprache ihre Erlebnisse bis zu einem gewissen Grade miteinander vergleichen. Dabei zeigt sich, daß gewisse sinnliche Erlebnisse verschiedener Menschen einander entsprechen, während bei anderen ein solches Entsprechen nicht festgestellt werden kann. Jenen sinnlichen Erlebnissen verschiedener Individuen, welche einander entsprechen und demnach in gewissem Sinne überpersönlich sind, wird eine Realität gedanklich zugeordnet. Von ihr, daher mittelbar von der Gesamtheit jener Erlebnisse, handeln die Naturwissenschaften, speziell auch deren elementarste, die Physik. Relativ konstanten Erlebnis-komplexen solcher Art entspricht der Begriff des physikalischen Körpers, speziell auch des festen Körpers."}
\end{quote}
Admittedly he only speaks of the ``sensory experiences of different people" and not explicitly of the transformations that convert sensations between them, nor of general covariance for that matter. Not explicitly, but almost: he eventually mentions physics; \emph{experiences} in physics can be called \emph{measurements}, and they tend to produce numbers; theory provides the transformations converting the numbers found by one person into those found by another. For measurements yielding a single number, the interpersonal `coincidence' at issue can be interpreted as numerical equality: only genuine \emph{scalars}---the same for everyone---would belong to the `superpersonal reality.' With measurements producing \emph{complexes} of numbers the notion of `coincidence' upon which reality rests is less straightforward: since numerical equality, for each component of the complex, would be much too strong, it will have to be a more holistic kind of correspondence, to do with the way the components change together. Vanishing is an important criterion: a complex whose components are \emph{wegtransformierbar} cannot be physically real---one whose components all vanish cannot `coincide' with one whose components don't. Of course the characteristic class of transformations is not the same in every theory; in general relativity it is the most general class (of transformations satisfying minimal requirements of continuity and differentiability). So it does not seem unreasonable to interpret the above passage as saying that \emph{only generally covariant notions represent reality in general relativity}.

Eight pages on Einstein speaks of geometry in a similar spirit:
\begin{quote}
In Euclidean geometry it is manifest that only (and all) quantities that can be expressed as invariants (with respect to linear orthogonal coordinates) have objective meaning (which does not depend on the particular choice of the Cartesian system). It is for this reason that the theory of invariants, which deals with the structural laws of invariants, is significant for analytic geometry.\footnote{``Offenbar haben in der euklidischen Geometrie nur solche (und alle solche) Größen eine objektive (von der besonderen Wahl des kartesischen Systems unabhängige) Bedeutung, welche sich durch eine Invariante (bezüglich linearer orthogonaler Koordinaten) ausdrücken lassen. Hierauf beruht es, daß die Invariantentheorie, welche sich mit den Strukturgesetzen der Invariante beschäftigt, für die analytische Geometrie von Bedeutung ist."}
\end{quote}
Here ``objective meaning" is explicitly attributed to invariance under the characteristic class of transformations.

In a letter to Paul Painlevé dated 7 December 1921 Einstein will be even more explicit, claiming that coordinates and quantities depending on them not only have no physical meaning, but do not even represent measurement results:
\begin{quote}
When one replaces $r$ with any function of $r$ in the $ds^2$ of the static spherically symmetric solution, one does not obtain a \emph{new} solution, for the quantity $r$ in itself has no physical meaning, meaning possessed only by the quantity $ds$ itself or rather by the network of all $ds$'s in the four-dimensional manifold. One always has to bear in mind that coordinates in themselves have no physical meaning, which means that they do not represent measurement results; only the results obtained by the elimination of coordinates can claim objective meaning.\footnote{\citet{Painlev}: ``\foreignlanguage{german}{Wenn man in der zentral-symmetrischen statischen Lösung für $ds^2$ statt $r$ irgend eine Funktion von $r$ einfügt, so erhält man keine \emph{neue} Lö[su]ng, da die Grösse $r$ an sich keinerlei physikalische Bedeutung hat, sondern nur die Grösse $ds$ selbst, oder besser gesagt das Netz aller $ds$ in der vierdimensionalen Mannigfaltigkeit. Es muss stets im Auge behalten werden, dass die Koordinaten an sich keine physikalische Bedeutung besitzen, das heisst, dass sie keine Messresultate darstellen, nur Ergebnisse, die durch Elimination der Koordinaten erlangt sind, können objektive Bedeutung beanspruchen. Die metrische Interpretation der Grösse $ds$ ist ferner keine "`pur imagination"', sondern der innerste Kern der ganzen Theorie. Die Sache verhält sich nämlich wie folgt: Gemäss der speziellen Relativitäts-Theorie sind die Koordinaten $x,y,z,t$ mittelst relativ zum Koordinaten-System ruhenden Uhren unmittelbar messbar, also hat auch die Invariante $ds$, definiert durch die Gleichung $ds^2=dt^2-dx^2-dy^2-dz^2$ die Bedeutung eines Messergebnisses.}"}
\end{quote}

The tension with the passages quoted in footnotes \ref{Gravitationswellen} and \ref{Energie} above is not without its significance for the relationist, who at this point can really question the legitimacy of a mathematical tolerance whose champion would develop an intransigence surprisingly reminiscent of the severity expressed by his previous opponents.

One can wonder what made Einstein change his mind, after Levi-Civita, Schrödinger and others had failed to persuade him. At the end of the foreword, dated 9 August 1920, to Cassirer's \emph{Zur Einstein'schen Relativitätstheorie} \citeyearpar{Cassirer} we discover that Einstein had read the manuscript and made comments. There he would have found the first thorough justification of the mathematical severity his opponents had expressed a few years before. We know how much the philosophical writings of Hume, Mach and Poincaré had influenced Einstein,\footnote{See \citet{Howard2005}.} and can conjecture that even here he was finally persuaded by a philosopher after the best mathematical physicists of the day had failed.

Be that as it may, it was too late to repent: the damage had been done, the (new) cause was already lost, and indeed the lenience Einstein promoted in 1918 continues to this day. General covariance\footnote{\emph{Cf}.\ \citet{Norton1993}.} is often disregarded or violated in general relativity: if a calculation works in one coordinate system, too bad if it doesn't in another; if energy conservation is upset by peculiar coordinates, never mind.

\subsection{Cassirer}\label{Cassirer}

\noindent Before going on we can briefly consider what Einstein would have found in Cassirer's manuscript.

Cassirer welcomed general relativity as confirming, even consolidating a philosophical and scientific tendency he had already described in \emph{Substanzbegriff und Funktionsbegriff} \citeyearpar{Cassirer1910}; a tendency that replaced the obvious things and substances filling the world of common sense, with abstract theoretical entities, relations and structures. Even the cruder objects of the naïve previous ontology derived their reality from `invariances' of sorts, but only apparent ones---mistakenly perceived by the roughness of our unassisted senses---which would be replaced by the more abstract and accurate invariants of modern theory.

Cassirer calls \emph{unity} ``the true goal of science."\footnote{\citet[p.\ 28]{Cassirer}: ``[die Einheit] ist das wahre Ziel der Wissenschaft. Von dieser Einheit aber hat der Physiker nicht zu fragen, \so{ob} sie ist, sondern lediglich \so{wie} sie ist -- d.~h.\ welches das Minimum der Voraussetzungen ist, die notwendig und hinreichend sind, eine eindeutige Darstellung der Gesamtheit der Erfahrungen und ihres systematischen Zusammenhangs zu liefern [\,\dots]."} It appears to have much to do with economy, of finding
\begin{quote}
a minimum of assumptions, which are necessary and sufficient to provide an unambiguous representation of experiences and their systematic context. To preserve, deepen and consolidate this unity, which seemed threatened by the tension between the principle of the constancy of the velocity of light, and the mechanical principle of relativity, the theory of relativity abandoned the uniqueness of measurement results for space and time quantities in different systems.\footnote{\emph{Ibid}.\ p.\ 28: ``Um diese Einheit, die durch den Widerstreit des Prinzips der Konstanz der Lichtgeschwindigkeit und des Relativitätsprinzips der Mechanik gefährdet schien, aufrecht zu erhalten und um sie tiefer und fester zu begründen, hat die Relativitätstheorie auf die Einerleiheit der Maßwerte für die Raum- und Zeitgrößen in den verschiedenen Systemen verzichtet."}
\end{quote}
Introducing differences where there were none before would seem rather to undermine or disrupt unity than to produce it \dots
\begin{quote}
But all these relativisations are so little in contradiction with the idea of the constancy and unity of nature, that they rather are required and carried out in the name of this very unity. The variation of space and time measurements represents the necessary \emph{condition}, through which the new invariants of the theory are first found and established.\footnote{\emph{Ibid}.\ p.\ 29: ``Aber alle diese Relativierungen stehen so wenig im Widerspruch zum Gedanken der Konstanz und der Einheit der Natur, daß sie vielmehr im Namen eben dieser Einheit gefordert und durchgeführt werden. Die Variation der Raum- und Zeitmaße bildet die notwendige \so{Bedingung}, vermöge deren die neuen Invarianten der Theorie sich erst finden und begründen lassen."}
\end{quote}
The foremost invariance is what we would typically call general covariance---which Cassirer considers ``the fundamental principle of general relativity":\footnote{\emph{Ibid}.\ p.\ 39: ``\foreignlanguage{german}{den Grundsatz der allgemeinen Relativitätstheorie, daß die allgemeinen Naturgesetze bei ganz beliebigen Transformationen der Raum-Zeit-Variablen ihre Form nicht ändern [\,\dots]}."}
\begin{quote}
Above all there is the general \emph{form} itself of the laws of nature, in which we must henceforth recognise the true invariant and as such the true logical basis of nature.\footnote{\emph{Ibid}.\ p.\ 29: ``Vor allem aber ist es die allgemeine \so{Form} der Naturgesetze selbst, in der wir nunmehr das eigentlich Invariante und somit das eigentliche logische Grundgerüst der Natur überhaupt zu erkennen haben."}
\end{quote}
Again, Cassirer sees Einstein's theory as a fundamental step in the transition between a common sense world made of (apparently invariant) `things,' to a more abstract and theoretical world of generally invariant mathematical objects, laws and relations.\footnote{\emph{ibid}.\ pp.\ 34-5} Only relations that hold for \emph{all} observers are genuinely objective,\footnote{\emph{Ibid}.\ p.\ 35: ``Wahrhaft objektiv können nur diejenigen Beziehungen und diejenigen besonderen Größenwerte heißen, die dieser kritischen Prüfung standhalten -- d.~h.\ die sich nicht nur für \so{ein} System, sondern für alle Systeme bewähren."} they alone can be objectively real ``natural laws."
\begin{quote}
We should only apply the term ``natural laws," and attribute objective reality, to relationships whose form does not depend on the peculiarity of our empirical measurement, on the special choice of the four variables $x_1,x_2,x_3,x_4$ which express the space and time parameters.\footnote{\emph{Ibid}.\ p.\ 39: ``Wir dürfen eben nur diejenigen Beziehungen Naturgesetze \so{nennen}, d.~h.\ ihnen objektive Allgemeinheit zusprechen, deren Gestalt von der Besonderheit unserer empirischen Messung, von der speziellen Wahl der vier Veränderlichen $\mathbf{x}_1\,\mathbf{x}_2\,\mathbf{x}_3\,\mathbf{x}_4$, die den Raum- und Zeitparameter ausdrücken, unabhängig ist."}
\end{quote}
Cassirer even associates \emph{truth} with general covariance:
\begin{quote}
The space and time measurements in each individual system remain relative: but the truth and generality of physical knowledge, which is nonetheless attainable, lies in the reciprocal correspondence of all these measurements, which transform according to specific rules.\footnote{\emph{Ibid}.\ p.~36: ``Die Raum- und Zeitmaße in jedem einzelnen System bleiben relativ: aber die Wahrheit und Allgemeinheit, die der physikalischen Erkenntnis nichtsdestoweniger erreichbar ist, besteht darin, daß alle diese Maße sich wechselseitig entsprechen und einander nach bestimmten Regeln zugeordnet sind."}
\end{quote}
Truth is not captured by a single perspective:
\begin{quote}
For relativity theory does not teach that whatever appears is real, but on the contrary warns against taking appearances which only hold with respect to a single system as scientific truth, in other words as an expression of the comprehensive and final legality of experience.\footnote{\emph{Ibid}.\ p.~50: ``Denn nicht, das jedem wahr sei, was ihm erscheint, will die [\,\dots] Relativitätstheorie lehren, sondern umgekehrt warnt sie davon, Erscheinungen, die nur von einem einzelnen bestimmten System aus gelten, schon für Wahrheit im Sinne der Wissenschaft, d.~h.\ für einen Ausdruck der umfassenden und endgültigen Gesetzlichkeit der Erfahrung zu nehmen."}
\end{quote}
Nor is it fully captured by an incomplete collection of perspectives; nothing short of \emph{all of them} will give the whole truth:
\begin{quote}
This will not be reached and ensured with respect to observations and measurements with respect to a single system, nor even with respect to arbitrarily many systems, but only through the reciprocal correspondences between results obtained in \emph{all} possible systems.\footnote{\emph{Ibid}.\ p.~50: ``Dieser wird weder durch die Beobachtungen und Messungen eines Einzelsystems, noch selbst durch diejenigen beliebig vieler solcher Systeme, sondern nur durch die wechselseitige Zuordnung der Ergebnisse \so{aller} möglichen Systeme erreicht und gewährleistet."}
\end{quote}
The point being that anything less than \emph{general} covariance isn't good enough: $U^{\mu}_{\nu}$, $t^{\mu}_{\nu}$ and $\mathit{\Gamma}^{\mu}_{\nu\kappa}$ are `linearly' covariant, in the sense that they behave like tensors with respect to linear transformations; but
\begin{quote}
Measurement in \emph{one} system, or even in an unlimited plurality of `privileged' systems of some sort, would yield only peculiarities in the end, rather than the real `synthetic unity' of the object.\footnote{\emph{Ibid}.\ p.\ 37: ``\foreignlanguage{german}{Die Messung in \so{einem} System, oder selbst in einer unbeschränkten Vielheit irgendwelcher "`berechtigter"' Systeme, würde schließlich immer nur Einzelheiten, nicht aber die echte "`synthetische Einheit"' des Gegenstandes ergeben}."}
\end{quote}
And ``overcoming the anthropomorphism of the natural sensory world view is," for Cassirer, ``the true task of physical knowledge," whose accomplishment is advanced by general covariance.\footnote{\emph{Ibid}.\ p.\ 37: ``Der Anthropomorphismus des natürlichen sinnlichen Weltbildes, dessen Überwindung die eigentliche Aufgabe der physikalischen Erkenntnis ist, wird hier abermals um einen Schritt weiter zurückgedrängt."} \citet[pp.~457-8]{Earman2006} is ``leery of an attempt to use an appeal to intuitions about what is physically meaningful to establish, independently of the details of particular theories, a general thesis about what can count as a general physical quantity"; we have seen that Cassirer was less leery, and so---as Earman is suggesting---was Einstein \dots

\subsection{Consistency}\label{consistency}
\noindent One hesitates---with or without Cassirer---to attach objective reality or even importance to things overly shaped by the peculiarities, point of view, state of motion or tastes of the subject or observer. Allowing him \emph{no} participation would be somewhat drastic, leaving at most the meagrest `truly objective' residue; but too much could make the object rather `unobjective,' and belong more to the observer than to the common reality. Appropriate transformation properties allow a moderate and regulated participation.

Is there an easy way of characterising how much participation would be too much? Of determining the `appropriateness' of transformation properties? Again: vanishing, annihilation seems an important criterion, as to which the relationist can demand agreement for physical significance; he will deny the reality of a quantity that can be transformed away, that disappears for some observers but not others.

But perhaps there is more at issue than just opinion or perspective. Much as one can wonder whether the different witnesses in Rashomon are \emph{lying}, rather than expressing reasonable differences in perspective; whether their versions are \emph{incompatible}, not just coloured by stance and prejudice---here the relationist may even complain about something as strong as \emph{inconsistency}, while his opponent sees no more than rival points of view.

Of an object that's at rest in one system but not in another\footnote{Observer $\mathit{\Xi}$ with four-velocity $V$ attributes speed $w=\sqrt{|g\textrm{(}\mathbf{w},\mathbf{w}\textrm{)}|}$ to body $\beta$ with four-velocity $W$,
where the (spacelike) three-velocity $\mathbf{w}$ is the projection
$$P_{V^{\negthinspace\perp}}W=\sum_{i=1}^3\langle dx^i,W\rangle\partial_i=W-g(V,W)V$$
onto the three-dimensional simultaneity subspace $V^{\negthinspace\perp}=\textrm{span}\{\partial_1,\partial_2,\partial_3\}$ orthogonal to $V$; and the projector $P_{V^{\negthinspace\perp}}=\langle dx^i,\cdot\,\rangle\partial_i$ is the identity minus the projector $P_V=g\textrm{(}V,\cdot\,\textrm{)}V$ onto the ray determined by $V$. Another observer
$\mathit{\Xi}'$ moving at $V'$ sees speed $w'=\sqrt{|g\textrm{(}\mathbf{w}',\mathbf{w}'\textrm{)}|}=\|P_{{V'}^{\negthinspace\perp}}W\|$ (all of this around the same event). Here we're supposing that one of the speeds vanishes.} one can say that \emph{it's moving \& isn't}, which sounds contradictory. Consistency can of course be restored with longer statements specifying perspective, but the tension between the short statements is not without significance---if the number were a scalar even they would agree. Similar considerations apply, \emph{mutatis mutandis}, to covariance; one would then speak of form or syntax being the same, rather than of numerical equality.

Consistency and reality are not unrelated. Consistency is certainly bound up with mathematical existence, for which it has long been considered necessary---perhaps even sufficient.\footnote{See \citet[p.~59]{Poincare}.} And in mathematical physics, how can the physical significance of a mathematical structure not be compromised by its inconsistency? If inconsistency prevents part of a formalism from `existing,' how can it represent reality? The relationist will argue that an object, like $t^{\mu}_{\nu}$, whose existence is complicated---perhaps even compromised---by an `inconsistency' of sorts (it's there, and it isn't), cannot be physically meaningful.

\subsection{The generation of gravitational waves}\label{binarystar}
\par\medskip
\begin{quote}
\emph{\small{LEX I. \emph{[\,\dots]} Majora autem planetarum et cometarum corpora motus suos et progressivos et circulares in spatiis minus resistentibus factos conservant diutius.}}
\end{quote}
\par\smallskip
\noindent We can now turn from the reality of gravitational waves to their very generation, about which the relationist can also wonder.

Belief in gravitational radiation rests largely on the binary star PSR 1913$\thinspace+$16, which loses kinetic energy as it spirals inwards (with respect to popular coordinates at any rate). If the kinetic energy is not to disappear without trace, it has to be converted, presumably into radiation. Since its disappearance is only ruled out by the conservation law, however, the very generation of gravitational waves must be subject to the perplexities surrounding conservation.\footnote{\emph{Cf}.\ \citet{Hoefer2}, \citet{Baker}.} If the conservation law is suspicious enough to make us wonder whether the lost energy is really radiated into the gravitational field, why take the polarisation of that radiation---which stands in the way of the full determination of inertia---seriously? As we were wondering in \S\ref{R&I}, couldn't it be no more than a purely decorative freedom, without reality or physical meaning? The binary star's behaviour and emission of gravitational waves can admittedly be calculated with great accuracy, but the calculations are not \emph{generally} covariant and only work in certain coordinate systems.

Even the `spiral' behaviour, associated so intimately with the loss of kinetic energy, is \emph{wegtransformierbar}. At every point along the worldlines $\sigma^1$ and $\sigma^2$ of the pulsars one can always choose (\emph{cf}.\ Joshua x, 13: ``the sun stood still, and the moon stayed") a basis $\mathbf{e}_{\mu}^r$ whose timelike vector $\mathbf{e}_0^r$ coincides with the four-velocity $\dot{\sigma}^r$ ($r=1,2$). Since nothing prevents the bases from being holonomic we can view them as natural bases $\mathbf{e}_{\mu}^r=\partial_{\mu}^r$ of a coordinate system, with respect to which $\dot{\sigma}^r$ will have components ($1,0,0,0$)---the three naughts being the components of the vanishing three-velocity $\mathbf{v}$. The coordinate system can be chosen so as to leave the pulsars at, say, the constant positions ($t,1,0,0$) and ($t,0,0,0$). If the pulsars don't move, if they have no `kinesis,' how can they lose a kinetic energy (which is after all a quadratic function of the three-velocity $\mathbf{v}$) they never had in the first place?\footnote{The pulsars are a bit large for low-dimensional idealisation (see \S\ref{midwife}); but one can still transform away the motions of representative worldlines---perhaps described by the centres of gravity---selected from their worldtubes. \emph{Cf}.\ \citet[p.\ 198]{WeylMuK}.}

It may be felt that the pulsars have a genuine angular momentum, with the right transformation properties; that they \emph{really are going around}. But angular momentum is about as coordinate-dependent as quantities get---its transformation properties could hardly be worse. The range of substitutions on which general relativity was built allows us to choose a coordinate system that eliminates the rotation by turning with the pulsars. If one feels instinctively that the rotation is real and legitimate, that it transcends coordinates, one's instincts are surreptitiously appealing---comparing the motion---to a background that \emph{general} relativity was conceived to do away with (but since seems to have found its way back). We are not really saying that such a backdrop is necessarily wrong or absent or unphysical or absurd, only that it should not be appealed to in general relativity, which was invented to get rid of it; the point we are making is more theoretical, conceptual and mathematical than physical. Certain coordinate systems may seem artificial, pathological, even perverse; but \emph{general} relativity is precisely the theory of such perversions, or rather of a generality encompassing so much that many surprising substitutions are admitted along with more mundane ones. Some transformations may savour of dishonest trickery; it might seem we are unscrupulously taking advantage of the full range of possibilities offered by general relativity, of substitutions lying on the fringe of legitimacy, out on the dark edges of a class too enormous to take seriously in its entirety. We can only repeat that the point of \emph{general} relativity is precisely its \emph{full generality}. Abstract talk of diffeomorphisms may make certain radical transformations less alarming---for are some diffeomorphisms more legitimate than others? Isn't general relativity the egalitarian theory putting them all on an equal footing?

Suppose we try to resolve the inward spiral into a rotation and a simple `inward,' `centripetal' motion. What about the centripetal motion? All that's needed for its elimination is a continuous recalibration of coordinates (leaving their directions unaffected), a time-dependent version of the transformation going from, say, inches to metres. In the next section we will appeal to \emph{geodesic deviation} to express the relationship between \emph{neighbouring} worldlines; but the pulsars are much too far apart for the construction of a \emph{well-behaved}, \emph{tensorial} acceleration of one pulsar with respect to the other.\footnote{Affine structure allows the (unambiguous) comparison of \emph{neighbouring}, not distant, directions.}

To question the reality or generation of gravitational waves, the relationist would demand general covariance---one of the central \emph{principles} of general relativity---\emph{as a matter of principle}, whereas his opponent will fall back on the more tolerant day-to-day pragmatism of the practising, calculating, approximating physicist, who views the theory more as an instrumental collection of recipes, perturbation methods, tricks and expedients, by which even the most sacred principles can be circumvented, than as a handful of fundamental and inviolable axioms from which all is to be deduced. General covariance may have been indispensable at first (it seems a whole crowd of midwives was assembled for so demanding a birth), but surely general relativity has now outgrown it \dots

\subsection{A Doppler effect}\label{Doppler}

\noindent The absolutist will be doubly satisfied by the discovery of gravitational waves, which would not only reinforce his belief in the underdetermination of inertia, but even allow absolute motion, as we shall now see.

We began with Newton's efforts to sort out absolute and relative \emph{motus}, first took (certain occurrences of) \emph{motus} to mean \emph{acceleration}, and accordingly considered absolute acceleration; but are now in a position to countenance absolute motion more literally. The four ontic-tidal-gravitational observables of Lusanna \& Pauri may even give us absolute \emph{position}: an observer capable of measuring them would infer his absolute position from the ontic-tidal-gravitational peculiarities of the spot---and even an equally absolute \emph{motion} from the variation of those peculiarities. But their measurement is anything but trivial, as one gathers from \S 2.2 of \citet{LusannaPauri2}. The importance of metrology for their programme is clear: if the four ontic-tidal-gravitational observables are in fact unobservable, why bother with them? We avoid all the formidable intricacies of metrology, faced with such competence and courage by Lusanna~\&~Pauri, by proposing a \emph{Gedankenexperiment} that's as simple as it is impossible: Let us say that relative motion is motion referred to some\emph{thing}---where by `thing' we mean a material object that has mass whatever the state of motion of the observer (materiality, again, \emph{is not an opinion}). Otherwise motion will be \emph{absolute}. Suppose an empty flat universe is perturbed by (\ref{harmonic}). Changes in the frequency $\omega$ measured by a roving observer would indicate absolute motion, and allow a reconstruction, through $\omega=k_aV^a$, of the observer's absolute velocity $V^a$.

Is this undulating space-time absolute, substantival,\footnote{Newton never seems to use words resembling `substance' in reference to his absolute space, whereas the literature about it is full of them.} Newtonian? It is absolute to the extent that according to the criterion adopted it admits absolute motion. But its absoluteness precludes its substantival reification, which would make the motion relative to some\emph{thing} and hence not absolute. Newton, though no doubt approving on the whole, would disown it, for ``Spatium absolutum [\thinspace\ldots] semper manet similare et immobile," and our undulating space-time is neither `similar to itself' ($R^a_{bcd}$ oscillates, though $R_{bd}$ vanishes identically) nor immobile.

We may remember that Newton spoke of revealing absolute \emph{motus} through its causes and effects, through forces. Absolute \emph{motion} is precisely what our thought experiment would reveal, and through forces, just as Newton wanted: the forces, for instance, registered by a (most sensitive) dynamometer linking the masses whose varying tidal oscillations give rise to the described Doppler effect.

The absolutist will claim, then, that gravitational waves are so real they wiggle the detector, and in so doing reveal absolute motion. But wiggling, the relationist will object, is not generally covariant: it can be transformed away. Let us continue to suppose, for simplicity, that the masses (two are enough) making up the detector are in the middle of nowhere, and not on the surface of the earth---whose gravitational field is not the point here. In what sense do they wiggle? As with the binary star, we can find coordinate systems that leave them where they are, say at ($t,1,0,0$) and ($t,0,0,0$). Both masses describe geodesics; how can things wiggle if they neither accelerate\footnote{\emph{Cf}.\ \citet[p.~80]{Lusanna}: ``all realistic observers are accelerated," for unaccelerated observers would have to be too small to be realistic; but see \S\ref{midwife} above.} nor move? The absolutist will reply that each mass, despite moving inertially, accelerates absolutely with respect to the other, for the tensorial, generally covariant expression ${d^2\xi^a}/{d\tau^2}=R^a_{0c0}\xi^c$ representing geodesic deviation cannot be transformed away (where $\xi^a$ is the separation, with components $\xi^{\mu}=\langle dx^{\mu}_a,\xi^a\rangle$, and $\tau$ is the proper time of the mass to which the acceleration of the other is referred). This puts the relationist in something of a corner, \emph{mathematically}---from which he could only emerge \emph{experimentally} by pointing out that the acceleration in question, however tensorial and covariant, has yet to be \emph{measured}.

\section{Final remarks}
\noindent The reader may feel, perhaps uneasily, that these explorations have been\dots\,exactly that; that they lack the factious zeal that so often animates the literature, giving it colour and heat and sentiment. But the enthusiast remains free to take sides, without being discouraged by our hesitating ambivalence.

Having viewed general relativity as a reply to the absolute inertial structure of Newtonian mechanics---which acts on matter despite being unobservable, and does not even react to it---we have wondered about the extent to which the inertia of general relativity is determined by matter and thus overcomes the absoluteness it was responding to.\footnote{Again, the very fact that matter constrains inertia at all makes their relationship more balanced than before, as an anonymous referee has pointed out; but a full assessment of how good a response (to the absolute features of Newtonian mechanics) general relativity proved should nonetheless consider the details of the relationship.} We have chosen to concentrate on punctual determination, paying little attention to the holistic, field-theoretical constraint contributed by distant circumstances and stipulations. And at a point the matter tensor $T^a_b$  underdetermines inertia by ten degrees freedom, eight of which can be eliminated by suitable gauge choices. The remaining two represent the polarisation of gravitational waves, whose reality the relationist can contest by insisting on general covariance; for the generation and energy-momentum of gravitational waves can, in appropriate senses, be transformed away.\footnote{Again, only at a single point will \emph{Wegtransformierbarkeit} be logically tidy, depending on the object ($t^{\mu}_{\nu}$ or $\mathit{\Gamma}^{\mu}_{\nu\tau}$ or whatever) alone; over broader regions much else would have to be considered as well; one would be reduced to a long and complicated enumeration of cases: \emph{here it holds, there it doesn't, there it might if only} etc.} Their (long awaited) detection, which may at first seem just as \emph{wegtransformierbar}, would in fact be generally covariant.

So gravitational waves have an awkward status in general relativity: though not as mathematically sturdy as one might want them to be, they aren't so flimsy the relationist can do away with them without qualms. If gravitational waves could be legitimately dismissed as a fiction, the determination of inertia by matter would be rather complete; and general relativity could be viewed as a satisfactory response to the absolute features of Newtonian mechanics that bothered Einstein.

\citet[p. 227]{BelotEarman2001} write that ``It is no longer possible to cash out the disagreement in terms of the nature of absolute motion (absolute acceleration will be defined in terms of the four-dimensional geometrical structure that substantivalists and relationalists \emph{agree} about)." Relationists and absolutists---as we call them---may well agree that absolute motion, or rather inertia, is represented by affine structure; but disagree about the nature of its determination by matter: only a relationist would contest the physical significance of the mathematical underdetermination at issue here.

Questioning the reality of gravitational waves is neither orthodox nor usual; but their bad transformation behaviour, which does not seem entirely meaningless, is worth dwelling on. While we await convincing, unambiguous experimental evidence, our belief in gravitational waves will (or perhaps should) be bound up with our feelings about general covariance, about general intersubjective agreement.
\\

\noindent
We thank Silvio Bergia, Roberto Danese, Dennis Dieks, Mauro Dorato, John Earman, Vincenzo Fano, Paolo Freguglia, Pierluigi Graziani, Catia Grimani, Niccolò Guicciardini, Marc Lachièze-Rey, Liana Lomiento, Luca Lusanna, Giovanni Macchia, Antonio Masiello, John Norton, Marco Panza, Carlo Rovelli, Tom Ryckman, George Sparling and Nino Zanghì for many fruitful discussions; and anonymous referees for helpful suggestions and comments.

\end{document}